\documentclass[desactivate]{aa}  
\usepackage{graphicx}
\usepackage{txfonts}
\usepackage[version=4]{mhchem}
\usepackage{xcolor}
\usepackage{ulem}
\usepackage{booktabs}
\usepackage{hyperref}
\usepackage[figuresright]{rotating}
\usepackage{mathtools}
\usepackage{float}
\usepackage{lscape}
\usepackage{csquotes}

\titlerunning{The impact of attenuation on cosmic-ray chemistry. I.}
\authorrunning{Roy, Gaches and Tan}

\begin{document} 

\title{The impact of attenuation on cosmic-ray chemistry:}
\subtitle{I. Abundances and chemical calibrators in molecular clouds}

\author{{Arghyadeb Roy\inst{\ref{inst1},\ref{inst2},}\thanks{This research was part of the Chalmers Astrophysics and Space Sciences Summer program.}} \and Brandt A. L. Gaches {\inst{\ref{inst3}}} \and Jonathan C. Tan {\inst{\ref{inst1},\ref{inst4}}}}

\institute{
    Department of Space, Earth and Environment, Chalmers University of Technology, Gothenburg 412 96, Sweden
    \label{inst1}
    \and
    Department of Chemical Sciences, Indian Institute of Science Education and Research Kolkata, West Bengal 741246, India
    \label{inst2}
    \and    
    Faculty of Physics, University of Duisburg-Essen, Lotharstraße 1, 47057 Duisburg, Germany
    \label{inst3}
    \and
    Department of Astronomy, University of Virginia, 530 McCormick Road, Charlottesville, VA 22904, USA
    \label{inst4}}

% \date{Received XX ; accepted XX}
 
\abstract{The chemistry of shielded molecular gas is primarily driven by energetic, charged particles dubbed cosmic rays (CRs), in particular those with energies under 1 GeV.\, CRs ionize molecular hydrogen and helium, the latter of which contributes greatly to the destruction of molecules. CR ionization initiates a wide range of gas-phase chemistry, including pathways important for the so-called ``carbon cycle'', \ce{C+}/\ce{C}/\ce{CO}. Therefore, the CR ionization rate, $\zeta$, is fundamental in theoretical and observational astrochemistry. Although observational methods show a wide range of ionization rates---varying with the environment, especially decreasing into dense clouds---astrochemical models often assume a constant rate.  To address this limitation, we employ a post-processed gas-phase chemical model of a simulated dense molecular cloud that incorporates CR energy losses within the cloud. This approach allows us to investigate changes in abundance profiles of important chemical tracers and gas temperature. Furthermore, we analyze analytical calibrators for estimating $\zeta$ in dense molecular gas that are robust when tested against a full chemical network. Additionally, we provide improved estimations of the electron fraction in dense gas for better consistency with observational data and theoretical calibrations for UV-shielded regions.}


\maketitle
%
%-------------------------------------------------------------------

\section{Introduction}
Cosmic rays (CRs) are highly energetic particles that permeate the interstellar medium (ISM), playing a crucial role in both the chemistry and physics of the densest gas phases. Notably, CRs of astrochemical importance have energies $< 1$ GeV. In molecular clouds, where ultraviolet (UV) radiation is heavily attenuated, CRs become the primary source of ionization, particularly within the cold, dense regions \citep{dalgarno2006,Grenier2015,padovani2020,gaches2025c}. These high-energy particles are accelerated in shock environments throughout the Galaxy and initiate key chemical reactions by colliding with hydrogen molecules (H$_2$). This collision typically results in the ionization of H$_2$, producing H$_2^+$, whi/ch quickly reacts further to form the tri-hydrogen cation (H$_3^+$):
\begin{equation*}
\tag{R1} \label{R1}
    \rm{CR+H_2 \rightarrow H_2^+ +e^-+CR'}
\end{equation*}
\begin{equation*}
\tag{R2} \label{R2}
    \rm{H_2^+ + H_2 \rightarrow H_3^+ + H}
\end{equation*} 
This process not only sets the ionization state of dense matter, but also influences the dynamics of the gas, affecting phenomena such as ambipolar diffusion, a proposed mechanism for magnetic flux dissipation and gravitational collapse \cite{Mouschovias1976}.

H$_3^+$ is a pivotal ion in the ISM, driving a rich ion chemistry that leads to the formation of critical molecules such as CO, HCO$^+$, NH$_3$ and H$_2$O in the gas phase \citep{tielens2005,dalgarno2006,bayet2011,caselli2012,indriolo2013,Bialy2015}. Furthermore, the interaction of H$_3^+$ with deuterated hydrogen (HD) initiates the deuteration process, crucial for the chemical evolution of star-forming regions \citep{ceccarelli2014,Kong2015,Kong2016,Chia-Jung2021}. Deuterium chemistry becomes especially important in dense cores where H$_2$D$^+$ can act as a major charge reservoir \citep{Bovino2020,Sabatini2024}, whereas at lower densities its role is less pronounced. As CRs are central to the physical and chemical processes within these regions, understanding and measuring their ionization rate per hydrogen molecule ($\zeta_{\ce{H2}}$; hereafter only $\zeta$) has been a significant focus in astrochemistry. Thus, the cosmic-ray ionization rate (CRIR) is a fundamental parameter that influences not only the chemical pathways but also the heating of dense gas, making it a critical factor in the study of molecular clouds and the broader ISM \citep{dalgarno2006,bisbas2015,padovani&gaches2024}.

Since the 1970s, a number of studies have addressed the issue of determining $\zeta$ from common observables \citep{guelin1977,wootten1979}. H$_3^+$ works as a tracer in absorption towards strong infrared sources in the diffuse medium ($A_V\lesssim 1$ mag), and its comparatively simple kinetics can be solved to estimate $\zeta$. For example, \citet{mccall2003,indriolo2007,Indriolo2012a} have used this technique, with typical findings on the order of $\zeta = 10^{-16}$ s$^{-1}$. New studies on electron fraction and CR ionization rate in NGC 1333 \citep{Pineda2024} and in OMC-2 \& OMC-3 \citep{Socci2024} show a diverse range of CRIR inside the cloud. As low-energy CRs move through molecular gas, they quickly lose energy due to ionizations, pion generation, and coulomb interactions \citep{Schlickeiser2002,Padovani2009}.

However, even though the CRIR gradient is crucial, it is frequently handled simply on the presumption of a single constant rate in astrochemical models. The attenuation of the CRIR as a function of the hydrogen-nuclei column density, also known as $\zeta(N)$, has been parameterized in a number of ways \citep[ are a few examples]{Padovani2009,Padovani2018,Morlino2015,Schlickeiser2016,Phan2018,Ivlev2018,Silsbee2019}. 

A widely adopted method of including the CRIR gradient involves prescribing $\zeta(N)$ analytically or through tabulated data \citep[e.g.,][]{Rimmer2012,Redaelli2021,Gaches2022a, Gaches2022b}. The most commonly applied is the attenuation models from \cite{Padovani2018}, which have been integrated into the public astrochemical tools {\sc Uclchem} \citep{O'Donoghue2022} and {\sc 3d-pdr} \citep{Gaches2019, Gaches2022a}. We note that \cite{Padovani2022} later updated the attenuation curves for both $\mathcal{L}$ and $\mathcal{H}$ models, leading to differences of a factor $\approx 20-30\%$; however, a polynomial fit was not presented in that work, and so the implementation in {\sc 3d-pdr} is still the \cite{Padovani2018} fits. Full numerical solutions of the CR transport equation into hydrodynamic simulations have been performed, with very limited chemical networks \citep[e.g.,][]{Rathjen2021}, but these are still highly constrained by the chemical network size. Further refinement has been achieved via one-dimensional spectrum-resolved CR transport models coupled with chemistry \citep[e.g.,][]{Gaches2019,Owen2021}. \cite{Gaches2022a} presented a significant advancement through the integration of a $\zeta(N)$ attenuation curve into the public {\sc 3d-pdr} code, allowing fully three-dimensional astrochemical modeling with CR gradients. This was followed by \cite{Redaelli2024}, with post-processing of three-dimensional magneto-hydrodynamic (3D MHD) simulations of prestellar cores using a spatially varying CRIR based on the density-dependent attenuation function from \cite{Ivlev2019}. More recently, \cite{Latrille2025} introduced a numerical framework for dynamically simulating CR propagation within MHD simulations, providing a self-consistent, multidimensional treatment of CR attenuation.

CO, one of the most abundant species in the dense ISM after \ce{H2} and \ce{He}, is found in regions shielded from UV radiation but still influenced by CRs. Its abundance is sensitive to CR-driven destruction via reactions with \ce{He+} ions, as described by Reaction 13 (\hyperref[tab:rate_coeff]{R13}), making it a key indicator of the balance between CR ionization and molecular cloud chemistry. 
\ce{HCO+}, on the other hand, is a direct product of CR ionization processes. CRs ionize \ce{H2}, leading to the formation of \ce{H3+} (\ref{R1} $-$ \ref{R2}), which reacts with CO to form \ce{HCO+} \citep[see Figure 8 of][]{Panessa2023}. As a result, \ce{HCO+} abundance is highly sensitive to the local CRIR, making it an excellent probe for estimating ionization rates in various environments. 
\ce{N2H+} is also significantly influenced by \ce{CO}, since \ce{CO} acts as its primary destroyer. In dense gas, CRs warm the environment, leading to \ce{CO} desorption into the gas phase, which in turn reduces the abundance of \ce{N2H+}. This interplay makes the \ce{N2H+}/\ce{HCO+} ratio a valuable tracer in CRIR studies. \ce{N2H+} forms primarily through the reaction of \ce{N2} with \ce{H3+} (\hyperref[tab:rate_coeff]{R18}) and remains abundant in regions where CO is depleted. This makes \ce{N2H+} particularly valuable in probing environments where CO depletion occurs. 

\ce{HCO+} and \ce{N2H+} have been widely used as diagnostic tools, as their relative abundances can reflect the state of \ce{CO} in dense gas, whether it is frozen onto grains or desorbed by thermal or non-thermal processes. Assuming similar abundances of \ce{CO} and \ce{N2} and comparable reaction rates for \ce{H3+} with \ce{N2} and \ce{CO}, the ratio \ce{N2H+}/\ce{HCO+} tends to approach $\sim1$. When \ce{CO} freezes out, \ce{N2H+} dominates, leading to \ce{N2H+}/\ce{HCO+} $\gg1$. Conversely, when \ce{CO} desorbs, it destroys \ce{N2H+}, converting it into \ce{HCO+}, resulting in \ce{N2H+}/\ce{HCO+} $\ll1$.

Together, \ce{H3+}, \ce{CO}, \ce{HCO+}, and \ce{N2H+} provide complementary insights into CRIRs, tracing different parts of molecular clouds based on their sensitivity to ionization, destruction, and formation processes. Their relative abundances, particularly the ratios of \ce{HCO+} to \ce{CO} and \ce{N2H+} to \ce{HCO+}, serve as effective diagnostics for understanding the effects of CRs on molecular cloud chemistry and physical conditions. This combination of tracers allows astronomers to map the ionization structure and chemical composition of CR-irradiated regions, shedding light on the influence of CRs on star formation, cloud evolution, and the ISM as a whole \citep{Indriolo2012a,MoralesOrtiz2014}.

This work aims to investigate the impact of CR ionization on the physics and chemistry in a dense molecular cloud and to compare the abundance profiles with the analytical models used to infer $\zeta$ from observable tracers and search for the potential applicability limits of these models. To achieve these goals, we post-processed a 3D MHD simulation of a molecular cloud with a photo-dissociation region (PDR) model that includes the $\zeta(N)$ prescription.
The abundance of the species is then fed into analytic chemical models to calculate the relative error between numerical and analytical methods.

\begin{table}[H]
\caption{{Polynomial coefficients, $c_k$ for Eq. (\ref{eq1}), from Table F.1
from \citet{Padovani2018}}
\label{tab:ck}}
\centering
\begin{tabular}{ccc}
    \hline
    \hline
{$k$} & {$\mathcal{L}$ Model} &{$\mathcal{H}$ Model} \\
    \hline
 0& $-3.331056497233 \times 10^{6}$   & $1.001098610761 \times 10^7$ \\
 1& $1.207744586503 \times 10^6$ & $-4.231294690194 \times 10^6$  \\
 2& $-1.913914106234 \times 10^5$ & $7.921914432011 \times 10^5$  \\
 3& $1.731822350618 \times 10^4$ & $-8.623677095423 \times 10^4$  \\
 4& $-9.790557206178 \times 10^2$ & $6.015889127529 \times 10^3$  \\
 5& $3.543830893824 \times 10^1$ & $-2.789238383353 \times 10^2$  \\
 6& $-8.034869454520 \times 10^{-1}$ & $8.595814402406 \times 10^0$  \\
 7& $1.048808593086 \times 10^{-2}$ & $-1.698029737474 \times 10^{-1}$  \\
 8& $-6.188760100997 \times 10^{-5}$ & $1.951179287567 \times 10^{-3}$  \\
 9& $3.122820990797 \times 10^{-8}$ & $-9.937499546711 \times 10^{-6}$  \\
\hline
\end{tabular}
\end{table}

The paper is organised as follows. Sect. \ref{methods} illustrates the simulation setup. In Sect. \ref{Results}, we report our findings from the model analysis. Sect. \ref{discussion} presents new analytical expressions for CRIR calibration and electron fraction estimations. Finally, Sect. \ref{conclusion} provides our conclusions.

\section{Methodology} \label{methods}

We used the density and velocity distributions of the \enquote{dense} cloud model from \citet{Bisbas2021}, based on subregions extracted from the MHD simulations of \citet{wu2017}. The subregions are $L=16.64$ pc cubes with $128^3$ uniform cells each. The total mass of the cloud is $M_\text{tot} = 7.43 \times 10^4 M_\odot$ with mean H-nuclei number density $\langle n_\text{H} \rangle = M_\text{tot}/(L^3 \mu m_\text{H}) \sim 466\,\, \text{cm}^{-3}$; where $m_\text{H}$ is the mass of the hydrogen nucleus, and $\mu = 1.4$ is the mean particle mass. The observed mean column density of H-nuclei is $\langle N_\text{tot}\rangle = \langle n_\text{H}\rangle L = 2.4\times10^{22}\,\, \text{cm}^{-2}$.
 
We followed a similar recipe as stated in \cite{Gaches2022a} to run a chemical model using publicly available astrochemical code\footnote{\url{https://uclchem.github.io/3dpdr/}}  {\sc 3d-pdr} \citep{bisbas2012}; thus calculating its atomic and molecular abundances using $\chi(i)=n(i)/n_\text{H}$ for each species $i$, gas temperature and level populations. 
We use a subset of the UMIST 2012 chemical network \citep{McElroy2013}, consisting of 77 species and 1158 reactions, along with standard ISM abundances at solar metallicity ($n_\text{He}/n_\text{H} = 0.1$, $n_\text{C}/n_\text{H} = 10^{-4}$, $n_\text{O}/n_\text{H} = 3 \times 10^{-4}$). This reduced network was developed to model key nitrogen chemistry, including species such as \ce{N2H+} and \ce{NH3}, while maintaining simplicity (see Appendix \ref{network_benchmark} for benchmark studies). Notably, the network does not include CO freeze-out and grain-surface chemistry. We include a treatment of grain-assisted recombination for \ce{H+}, \ce{He+}, and \ce{C+}, which has been found to impact the chemistry of \ce{C+} and \ce{CO} and also the electron fraction \citep{Gong2017}. Similar to \citet{Bisbas2021} and \citet{Gaches2022a}, we employed an external isotropic far-ultraviolet (FUV) intensity of $G_0 = 10$ \citep[normalized as per][]{Draine1978}, a metallicity of $Z = 1 Z_\odot$, and a microturbulent dispersion velocity of $v_\text{turb} = 2$ km s$^{-1}$. This prescription enabled us to examine the effects of CR attenuation on more realistic self-generated clouds, despite the simulations being developed with specific external radiation fields. 

\begin{figure}[H]
    \centering
    \includegraphics[width=1.0\linewidth]{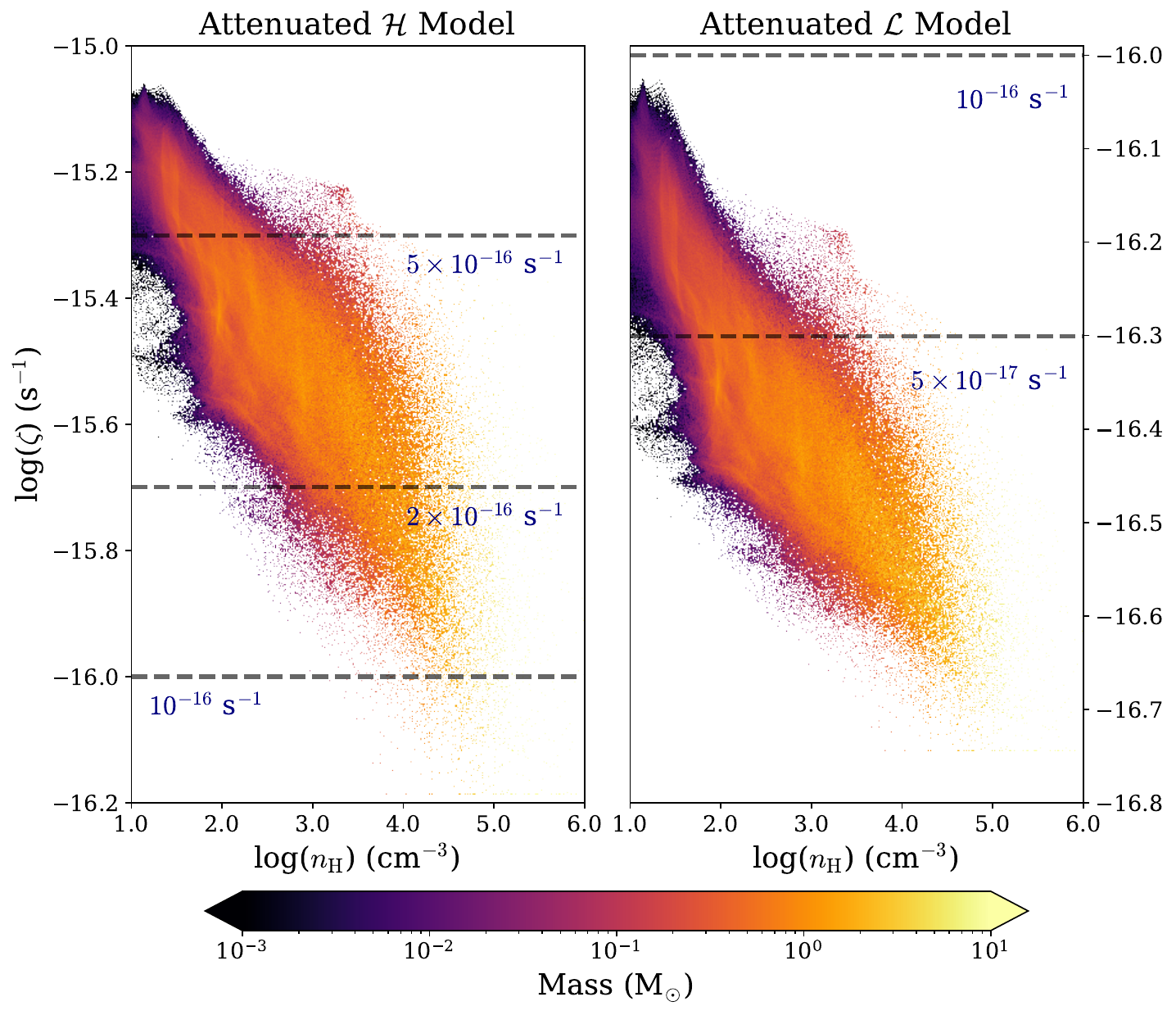}
    \caption{Mass weighted local CRIR---density 2D-PDF for the attenuated High $(\mathcal{H})$ and Low $(\mathcal{L})$ models applied in the \enquote{dense} cloud physical model (see text). The constant CRIR values are denoted as dashed lines.}
    \label{fig:zeta_nH}
\end{figure}

The CR attenuation is included using the polynomial relationship\footnote{{\sc 3d-pdr} can also solve the CR attenuation using a spectrum-resolved solver, but in this work, we used the prescribed polynomial function due to memory constraints and ability to compare with contemporary studies.} presented by \citet{Padovani2018} for CRIR versus hydrogen column density, $\zeta(N)$,
\begin{equation} \label{eq1}
    \text{log}_{10}\, \bigg(\frac{\zeta(N)}{\text{s}^{-1}}\bigg) = \sum_{k=0}^9 c_k \bigg[\text{log}_{10}\,\bigg(\frac{N_\text{H,eff}}{\text{cm}^{-2}}\bigg)\bigg]^k\,,
\end{equation}
where $N$ represents the total hydrogen column density \enquote{seen} by the CR and the coefficients, $c_k$, were obtained from Table F.1 of \citet{Padovani2018}, which is also displayed in Table \ref{tab:ck}. 
Although previous calculations in the literature \citep{Gaches2022a, Gaches2022b, Luo_2024} were solely based on the $\mathcal{H}$ model, proposed by \citet{Ivlev2015}, to replicate the diffuse gas CRIR estimates \citep[e.g.,][]{Indriolo2012a,Indriolo2015,neufeld2017}, recent constraints from \ce{C2} observations convey that the actual CRIR lies between the $\mathcal{H}$ and $\mathcal{L}$ models, rather than fully aligning with either extreme \citep{Neufeld2024,Obolentseva2024}. Hence, in this work, we consider both spectral models to comprehensively assess the impact of attenuation with the diffuse gas column density upper limit of $10^{19}$ {cm}$^{-2}$. We also run four constant-CR ionization models, $\zeta_c = (0.5, 1, 2, 5) \times 10^{-16}$ {s}$^{-1}$, to compare against our CR-attenuated models.

\section{Results} \label{Results}

\subsection{Physical properties}

The CRIR, $\zeta$, alone does not fully encompass the intricacies of CR-driven chemistry. The efficiency of ionization is contingent on the CR flux and influenced by local conditions such as hydrogen number density, $n_\text{H}$, and the electron density, $n_e$. The interplay between these parameters shapes the ionization-recombination balance, ultimately guiding the chemical evolution of interstellar clouds. The relationship between $\zeta$, and $n_\text{H}$ is particularly insightful. As depicted in Fig. \ref{fig:zeta_nH}, $\zeta$ decreases with increasing $n_\text{H}$ in both the CR-attenuated models alongside expressing the higher and lower limit of the CRIR in both models. This trend is expected because higher densities are associated with greater column densities, which in turn attenuate CR flux and limit its penetration depth, reducing the ionization rate. This behaviour holds true in regimes where the CR flux is isotropic, transport is dominated by energy losses rather than turbulent diffusion, and no embedded CR sources exist.

The $\zeta/n_\text{H}$ ratio represents the ionization rate per hydrogen molecule normalized by the hydrogen number density. This metric is particularly useful for comparing the strength and dominance of CR ionization across environments with varying densities. For instance, in diffuse molecular clouds (low $n_\text{H}$), $\zeta/n_\text{H}$ provides insight into the effectiveness of CR in driving ionization, similar to X-ray ionization characterization \citep{Wolfire2022}. This makes it a valuable parameter for understanding the impact of CRs in both diffuse and dense cloud environments.

The $\zeta/n_e$ (where, $\chi_e = n_e/n_\text{H}$) ratio, on the other hand, provides insights into the ionization-recombination balance, offering a direct measure of the ionization fraction. This parameter is particularly sensitive to the availability of free electrons, which play a critical role in determining the overall ionization state of the medium. While $\zeta/n_\text{H}$ captures the attenuation and strength of ionization relative to hydrogen density, $\zeta/n_e$ reflects the local ionization environment and the fraction of free electrons.

\begin{figure*}
    \centering
    \includegraphics[width=0.85\linewidth]{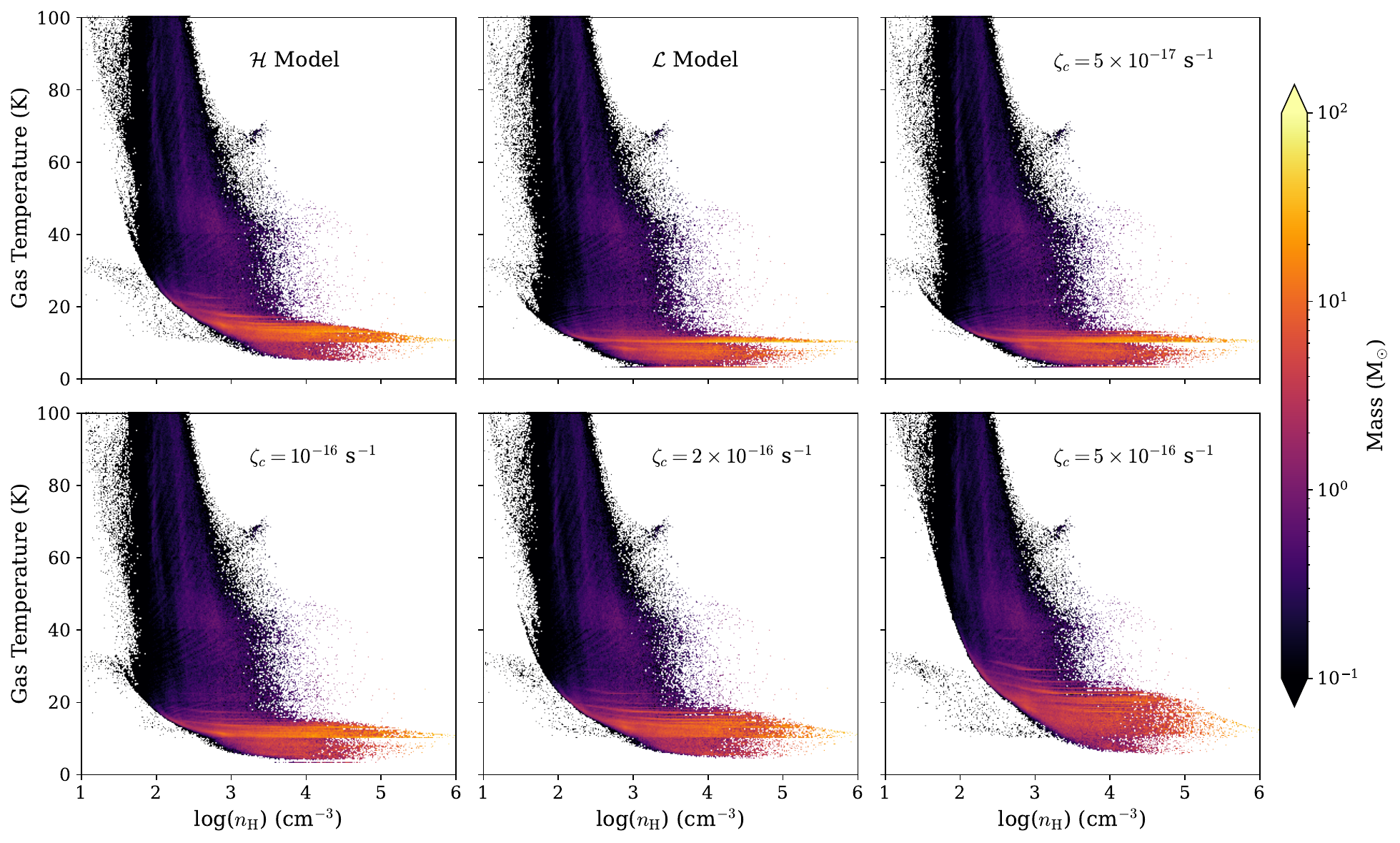}
    \caption{Mass-weighted density---temperature phase plot in semi-logarithmic scale (logarithmic in density). Density increases from left to right. The CR model is annotated in each panel.}
    \label{fig:gas_tem}
\end{figure*}

Fig.~\ref{fig:gas_tem} represents the mass-weighted density---temperature phase plot on a semi-logarithmic scale (logarithmic in density), highlighting the effect of CR heating. As evident from the distribution, the majority of the mass resides in cold, dense gas, with minimal CR-induced heating ($\sim 10$ K).

\subsection{Chemical abundances}

\begin{figure*}
    \centering
    \includegraphics[width=0.85\linewidth]{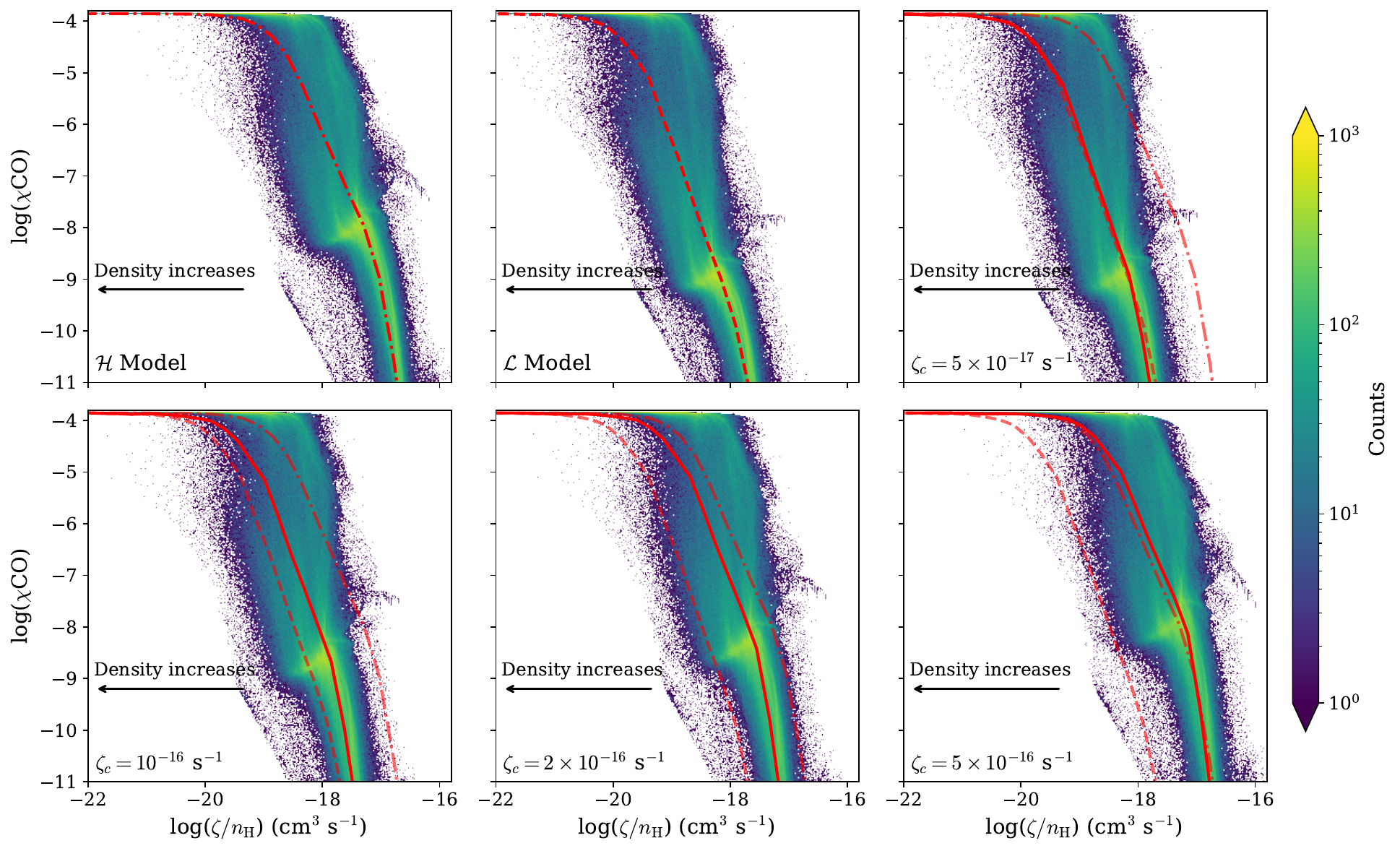}
    \caption{CO abundance profile, $\chi(\text{CO})$ versus effective ionization rate, $\zeta/n_\text{H}$. Density increases from right to left. The CR model is annotated in each panel. The red lines denote the binned averaged abundance profiles in log-log space. The dashed-dotted lines show the $\zeta(N)_\mathcal{H}$ model, and dashed lines show the $\zeta(N)_\mathcal{L}$ model abundance profile, for comparison.}
    \label{fig:CO_abund}
\end{figure*}

\begin{figure*}
    \centering
    \includegraphics[width=0.85\linewidth]{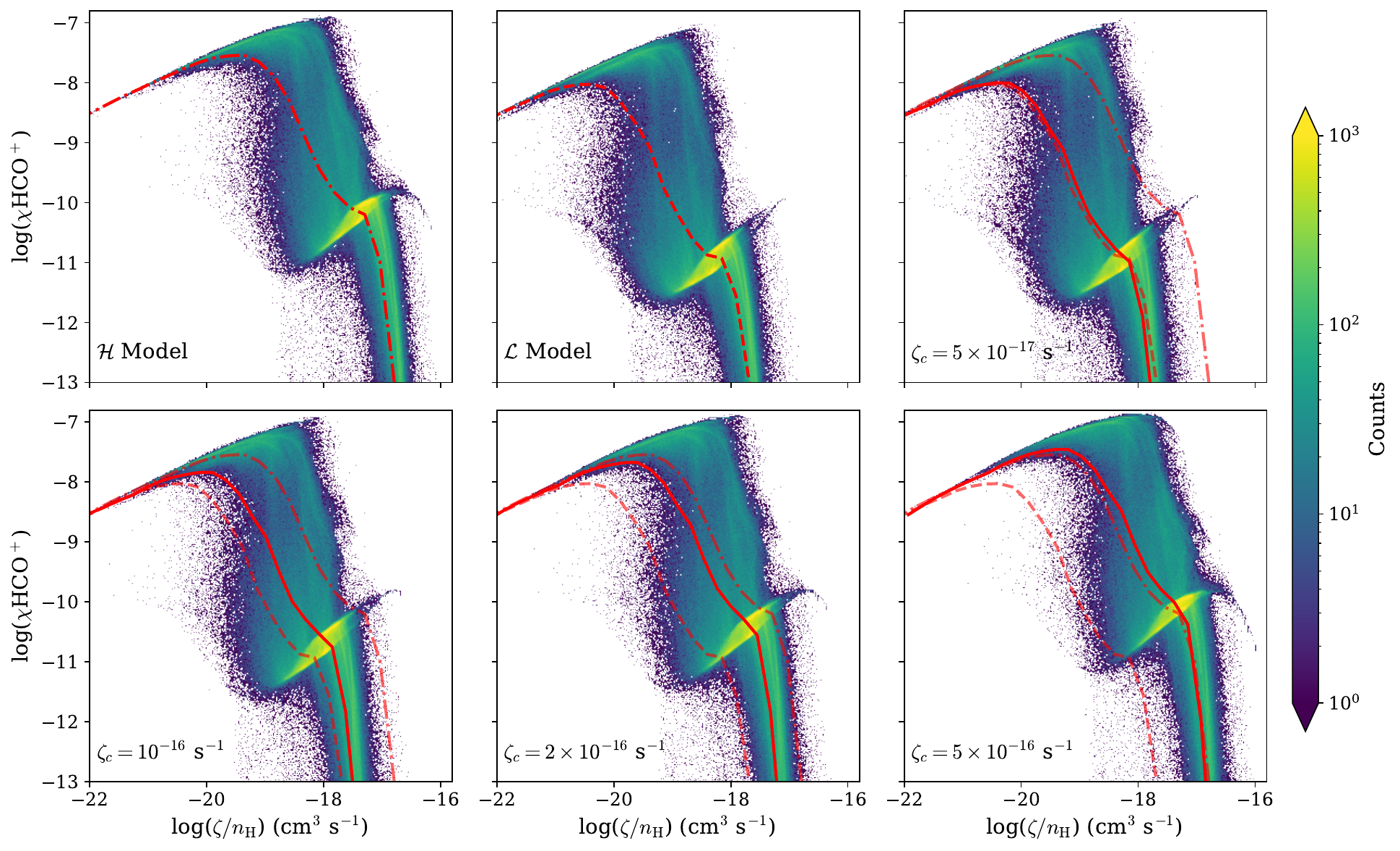}
    \caption{As in Fig. \ref{fig:CO_abund}, but for \ce{HCO+} abundance profile. Density increases from right to left.}
    \label{fig:HCOp_abund}
\end{figure*}

\begin{figure*}
    \centering
    \includegraphics[width=0.85\linewidth]{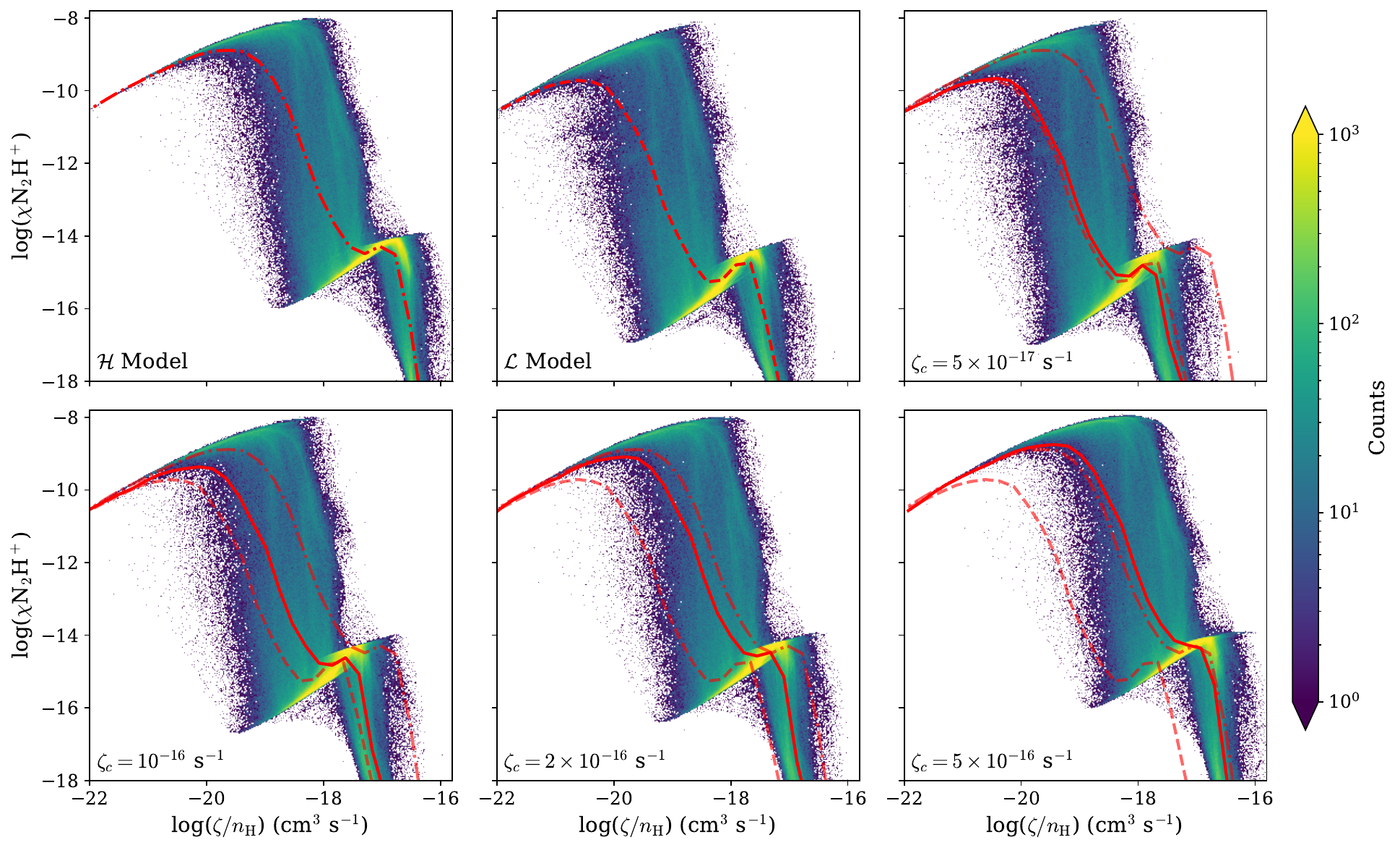}
    \caption{As in Fig. \ref{fig:CO_abund}, but for \ce{N2H+} abundance profile. Density increases from right to left.}
    \label{fig:N2Hp_abund}
\end{figure*}

\begin{figure*}
    \centering
    \includegraphics[width=0.85\linewidth]{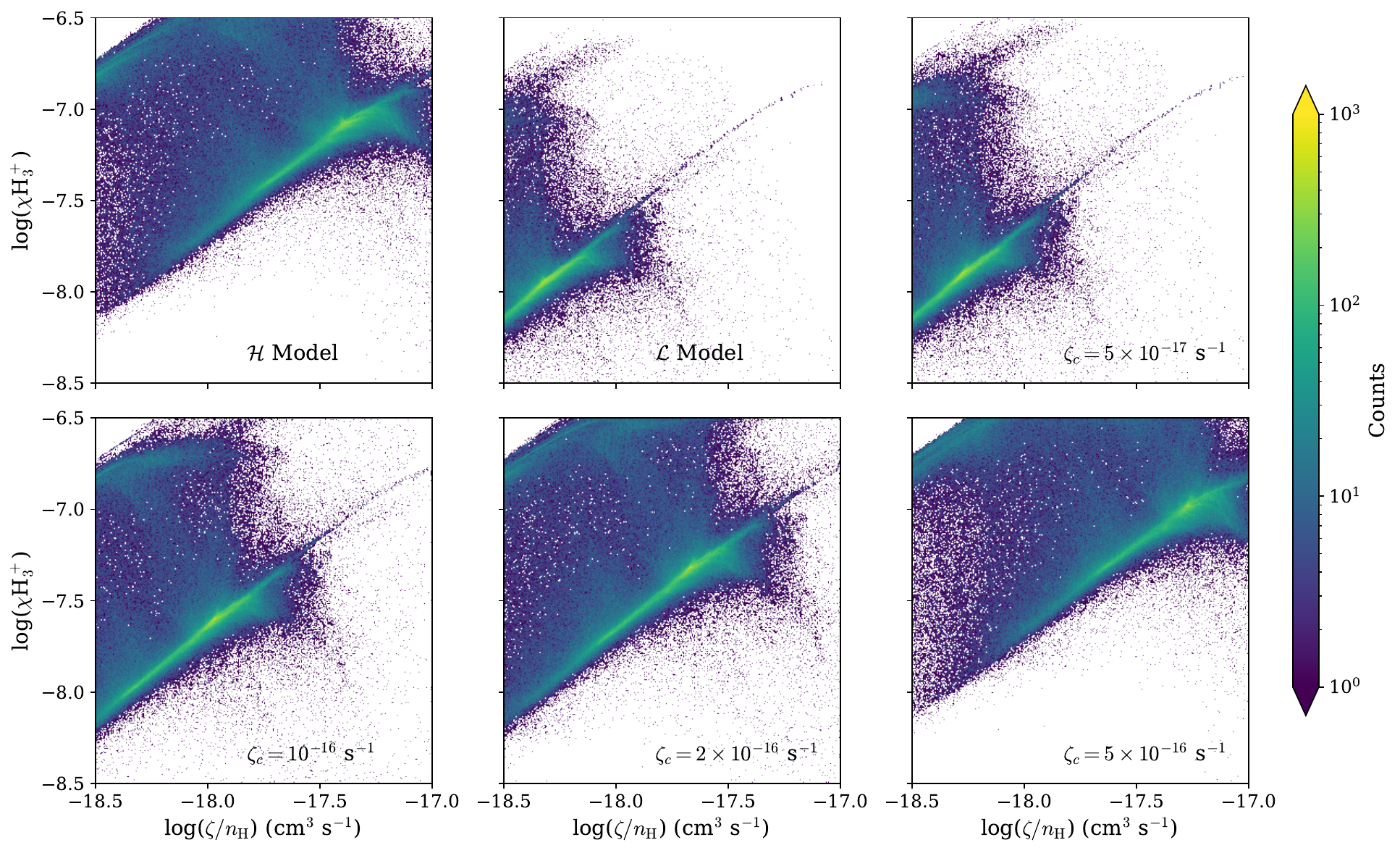}
    \caption{Zoomed ($-18.5 \leq \log(\zeta/n_\text{H}) \leq -17.0$) and filtered (\text{UV} $\leq 1 {G}_0$) version of the H$_3^+$ abundance profile.}
    \label{fig:H3+_zoomed}
\end{figure*}

The abundances of \ce{CO}, \ce{HCO+}, and \ce{N2H+} are plotted as functions of the effective ionization rate, $\zeta/n_\text{H}$, for all six ionization models, in Figs. \ref{fig:CO_abund} $-$ \ref{fig:N2Hp_abund}, respectively. The binned averages, in log-log space, are represented by red lines. The dashed-dotted lines show the abundance profile of the $\zeta(N)_\mathcal{H}$ model, and dashed lines show the abundance profile of the $\zeta(N)_\mathcal{L}$ model for comparison. The linear-relative error in the average abundance profiles for a particular constant CRIR model in comparison to the $\zeta(N)_\mathcal{H,L}$ model is calculated by,
\begin{equation} \label{eq:rel_Err}
   \varepsilon(\chi_i) = \dfrac{\chi_{i,\zeta_c} - \chi_{i,\zeta{(N)}_\mathcal{H,L}}}{\chi_{i,\zeta{(N)}_\mathcal{H,L}}}\,,
\end{equation}
and are displayed in Fig. \ref{fig:all_relerr} in the Appendix. 

The abundance profiles of \ce{CO}, \ce{HCO+}, and \ce{N2H+} vary significantly as functions of $\zeta/n_\text{H}$. In the low-density regime, where $\zeta/n_\text{H}$ is high, the abundance of \ce{CO} is greatly diminished. 
However, it is important to note that in diffuse gas, even in the absence of CRs, the UV radiation field plays a dominant role in driving the chemistry, significantly limiting the formation and survival of \ce{CO}. The scarcity of CO at low densities has a direct impact on the abundance of \ce{HCO+}, as \ce{CO} serves as its primary precursor. Consequently, \ce{HCO+} is extremely rare in this regime, with its abundance relative to CO being as low as $10^{-4}$. Similarly, \ce{N2H+} shows limited presence in this region, with its abundance approximately $10^{-5}$ times that of \ce{HCO+}. This limitation is due to the restricted availability of \ce{N2}, its precursor, and the intense competition between different ions for reactions.

In the intermediate density regime, where $\zeta/n_\text{H}$ begins to decrease, \ce{CO}, shielded from the UV field, starts to increase in abundance as the rates of its formation begin to counterbalance the rates of its destruction. CR attenuation also becomes more effective at protecting CO from CR-induced destruction, allowing it to accumulate. This increase in CO abundance marks a turning point for \ce{HCO+} formation. In this regime, the destruction of CO by \ce{H3+} produces intermediate species that drive the formation of \ce{HCO+}. As a result, the abundance of \ce{HCO+} rises sharply, reflecting the shift in chemical equilibrium. The increase in \ce{HCO+} abundance is accompanied by changes in the \ce{N2H+} profile. In the early part of the intermediate regime, \ce{N2H+} becomes more abundant in regions where CO has not yet accumulated to levels sufficient to act as a significant destroyer of \ce{N2H+}. However, as CO abundance continues to rise, it reacts with \ce{N2H+} (\hyperref[tab:rate_coeff]{R19}), gradually depleting the latter. 

At higher densities, the abundance of CO reaches a point where it becomes nearly independent of $\zeta$. In these dense environments, the rate of CO formation surpasses its destruction rate, resulting in saturation at the elemental carbon abundance $(\chi(\ce{CO})\approx A_\text{C})$; although this convergence occurs because CO freeze-out onto grain surfaces has not been considered in the network. This underscores the dominant role of \ce{CO} as the primary reservoir of carbon in these regions. The abundance of \ce{HCO+} continues to decrease in this regime through reactive collisions with electrons (\hyperref[tab:rate_coeff]{R12}), thus producing \ce{CO}.
A similar characteristic is observed for the abundance profile of \ce{N2H+}, where \ce{CO} collisions act as the primary destruction mechanism of \ce{N2H+}. 

The binned average abundance profiles indicate that the $\mathcal{L}$ model produces results very similar to a constant $\zeta_c = 5 \times 10^{-17}$~s$^{-1}$ model. As we consider higher constant CRIR models, the binned averages gradually shift from the reference line of the $\mathcal{L}$ model toward that of the $\mathcal{H}$ model. For the physical conditions considered here, the $\mathcal{L}$ model shows relatively little attenuation, whereas the $\mathcal{H}$ model exhibits factors of 5 or more in the abundance differences due to attenuation effects.

\section{Discussion} \label{discussion}
\subsection{Benchmarking \ce{H3+} analytic chemistry}

\begin{table*}
\caption{{Important reactions and their rate Coefficients, taken from UMIST2012 database.}
\label{tab:rate_coeff}}
\centering
\begin{tabular}{c c c}
    \hline
    \hline
{Reaction No.} & {Reaction} & {Rate Coefficient (cm$^3$ s$^{-1}$)} \\
    \hline
(R2) & $\rm{H_2^+ + H_2 \rightarrow H_3^+ + H}$ &  $k_2 = 2.08\times10^{-9}$ \\
(R3) & $\rm{H_2^+ + e^- \rightarrow H + H}$ &  $k_3 = 1.6\times10^{-8}(T/300)^{-0.43}$\\ 
(R4) & $\rm{H_2^+ + H \rightarrow H_2 + H^+}$ & $k_4 = 6.4\times10^{-10}$ \\
(R5)& $\rm{H_3^+ + e^- \rightarrow H_2 + H}$ & $k_5=2.34\times10^{-8}(T/300)^{-0.52}$\\
(R6)&$\rm{H_3^+ + e^- \rightarrow H + H + H}$ & $k_6 = 4.36\times10^{-8}(T/300)^{-0.52}$ \\
(R7)& $\rm{H_3^+ + CO \rightarrow HCO^+ + H_2}$ & $k_7=1.36\times10^{-9}(T/300)^{-0.14}\exp(3.4/T)$ \\
(R8)& $\rm{H_3^+ + CO \rightarrow HOC^+ + H_2}$ & $k_8=8.49\times10^{-10}(T/300)^{0.07}\exp(-5.2/T)$\\
(R9)& $\rm{H_3^+ + O \rightarrow OH^+ + H_2}$ &  $k_9 = 7.98\times10^{-10}(T/300)^{-0.16}\exp(-1.4/T)$\\
(R10)& $\rm{H_3^+ + O \rightarrow H_2O^+ + H}$ & $k_{10} = 3.42\times10^{-10}(T/300)^{-0.16}\exp(-1.4/T)$\\
(R11)&  $\rm{H_3^+ + C \rightarrow CH^+ + H_2}$ & $k_{11}=2.00\times10^{-9}$ \\
(R12) & $\rm{HCO^+ + e^- \rightarrow H + CO}$ & $k_{12}= 2.40 \times 10^{-7}(T/300)^{-0.69}$\\
(R13) & $\rm{He^+ + CO \rightarrow O+C^++He}$ & $k_{13} = 1.60\times10^{-9}$ \\
(R14) & $\rm{He^+ + O_2 \rightarrow O^+ + O +He}$ & $k_{14} = 1.10\times10^{-9}$\\
(R15) & $\rm{C^+ + e^- \rightarrow C + h\nu}$ & $k_{15} = 2.36\times10^{-12}(T/300)^{-0.29}\exp(17.6/T)$ \\
(R16) & $\rm{C^+ + O_2 \rightarrow CO^+ + O}$ & $k_{16} = 3.42\times 10^{-10}$\\
(R17) & $\rm{C^+ + O_2 \rightarrow CO + O^+}$ & $k_{17}=4.54\times10^{-10}$\\
(R18) & $\rm{H_3^+ + N_2 \rightarrow N_2H^+ + H_2}$ & $k_{18}=1.80\times10^{-9}$\\
(R19) & $\rm{CO + N_2H^+ \rightarrow HCO^+ + N_2}$ & $k_{19}=8.80\times10^{-10}$\\
\hline
\end{tabular}
\tablefoot{Numerical subscripts on rate coefficients correspond to the reaction numbers stated in the text. The \ce{H3+}$-$electron recombination rate coefficient ($k_e$) is equal to $k_5+k_6$. The rate coefficient for proton transfer to CO, denoted as $k_\text{CO}$, is calculated as $k_7 + k_8$. The rate coefficient for destruction via proton transfer to O, $k_\text{O}$, is equal to $k_9 + k_{10}$. The rate coefficient for \ce{O2} destruction via reaction with \ce{C+}, $k_{\text{C}^+}$, is equal to $k_{16}+k_{17}$.  Temperature-dependent coefficients are expressed in terms of gas kinetic temperature ($T$).}
\end{table*}

\begin{figure*}
    \centering
    \includegraphics[width=0.85\linewidth]{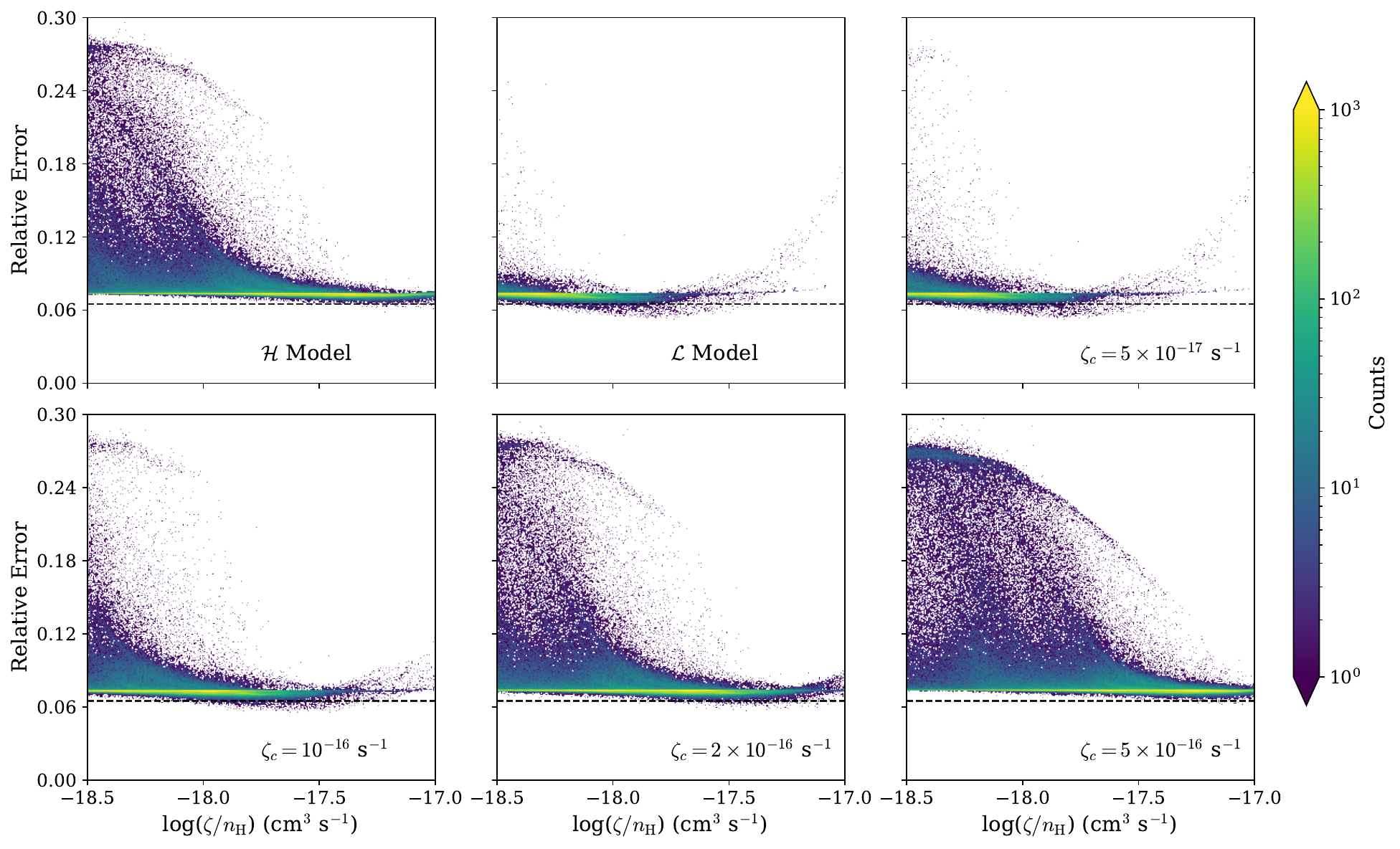}
    \caption{Relative error between the numerical abundance profile ({\sc 3d-pdr}) and the modified analytical expressions of H$_3^+$ (Eq. \ref{eq:2nd_order_mod}). The constant offset of 6.5\% is denoted by the black dashed line.}
    \label{fig:H3+_relerr_mod}
\end{figure*}

H$_3^+$ plays a critical role in interstellar chemistry, serving as an essential tracer of CR ionization. Due to its relatively simple structure and the fact that its formation and destruction pathways are well understood, H$_3^+$ can be used as a reliable probe to determine the CRIRs, especially in diffuse molecular clouds. Its abundance is determined by the ionization of H$_2$, making it an ideal candidate for analytically calibrating CR flux and ionization rates.

The $\chi(\text{H}_3^+) - \zeta/n_\text{H}$ phase space for our different simulations is shown in Fig. \ref{fig:H3+_abundace}. The abundance profile aligns with the expected dependence on CRIR. In low-density (or high-CRIR environments), a higher H$_3^+$ abundance is observed, reflecting the increased ionization activity. As the density increases, the H$_3^+$ abundance converges towards a certain upper limit, maintaining a relatively steady value throughout the intermediate-high density regime before dropping down at higher densities. A clear linear trend in the log-log phase space, between the region of $-18.5 \leq \log(\zeta/n_\text{H}) \leq -17.0$, is observed. With the expectation that this selected region will serve as a reliable reference for the analytical H$_3^+$ calibrators commonly used in the literature to constrain CR-ionization rates, we proceeded to further analyze the region. This analysis aims to validate the applicability of these calibrators under the observed conditions, providing a more accurate assessment of CR effects. A zoomed and filtered version of the selected region is displayed in Fig. \ref{fig:H3+_zoomed}.

\cite{Indriolo2012a} in their study used H$_3^+$ line of sight column density measurements to estimate CRIRs in diffuse clouds by comparing observed values with model analytic calibrators. Thus, we aim to test the robustness of these analytic calibrators using our numerical modeling results. The goal is to determine the extent to which the calibrations of H$_3^+$ match numerical simulations, providing a deeper understanding of how CRs drive chemistry in different interstellar regions and to which extent the calibrators can capture the CR chemistry. The following expressions of the calibrators are used in \cite{Indriolo2012a},
\begin{align*}
    \tag{3}
    \label{eq:1st_order}
    & \bigg(\frac{\zeta}{n_\text{H}}\bigg) = k_e\chi_e \frac{\chi(\text{H}_3^+)}{\chi(\text{H}_2)}
    \notag\\ %\quad\textit{and},\\
    & \bigg(\frac{\zeta}{n_\text{H}}\bigg)\, \chi(\text{H}_2) = \chi(\text{H}_2^+) [ k_3\chi_e  + k_4\chi (\text{H})  + k_2\chi (\text{H}_2) ] \,,\\
    & k_2\chi (\text{H}_2) \chi(\text{H}_2^+) = \chi(\text{H}_3^+) [k_e\chi_e  + k_{\text{CO}}\chi (\text{CO})  + k_9\chi (\text{O}) ] \,
    \tag{4}
    \label{eq:2nd_order}
\end{align*}

\begin{figure*}
    \centering
    \includegraphics[width=0.85\linewidth]{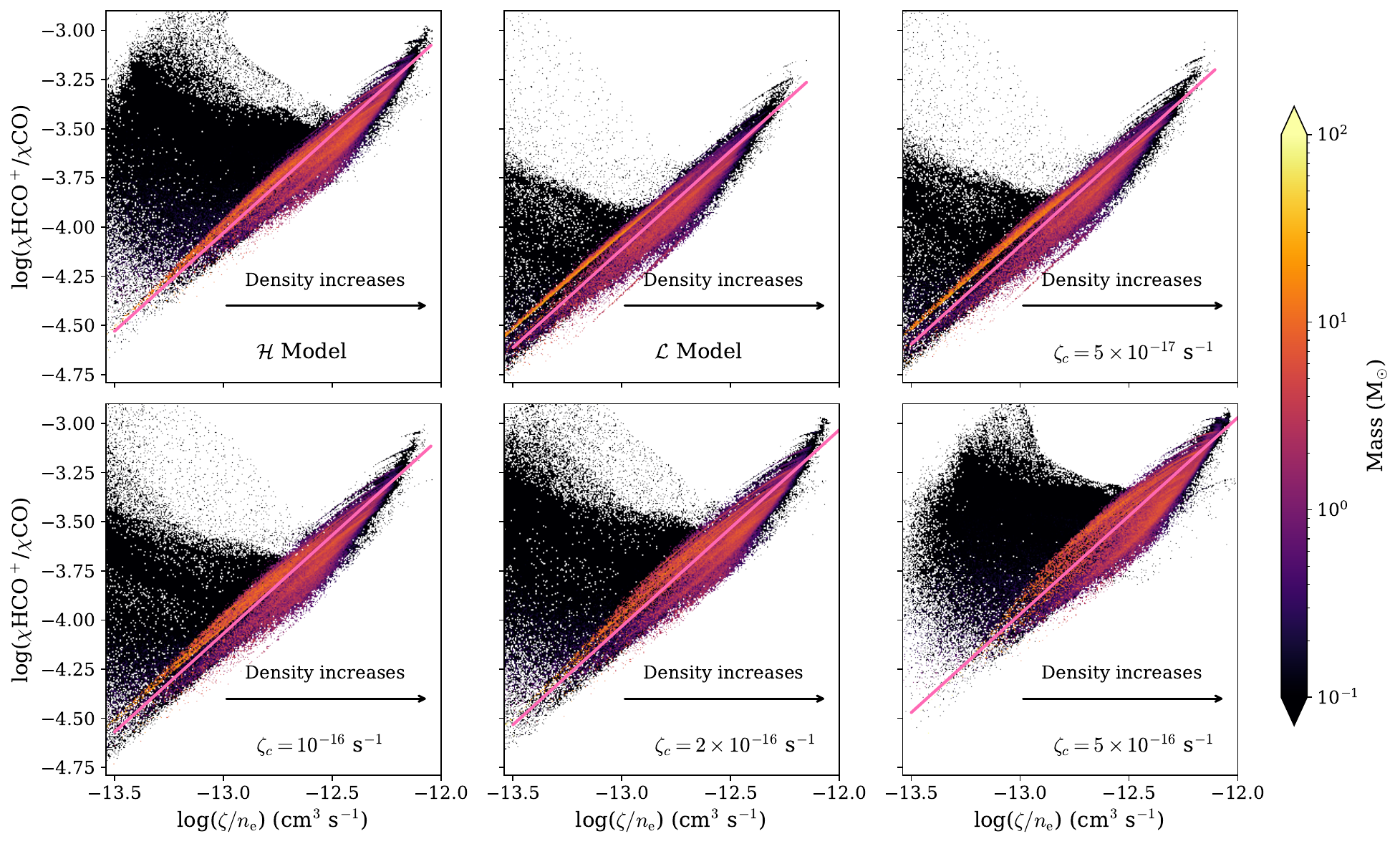}
    \caption{\ce{HCO+} and \ce{CO} abundance ratio, $\chi(\ce{HCO+})$/$\chi(\text{CO})$ versus $\zeta/n_e$. Density increases from left to right. The CR model is annotated in each panel. The data is filtered under the constraint: UV $<$ 0.1 $G_0$ and mass-weighted. A linear fit (pink line) in log–log space with slope $=1$ is overlaid.}
    \label{fig:HCO+/CO}
\end{figure*}

\begin{figure*}
    \centering
    \includegraphics[width=0.85\linewidth]{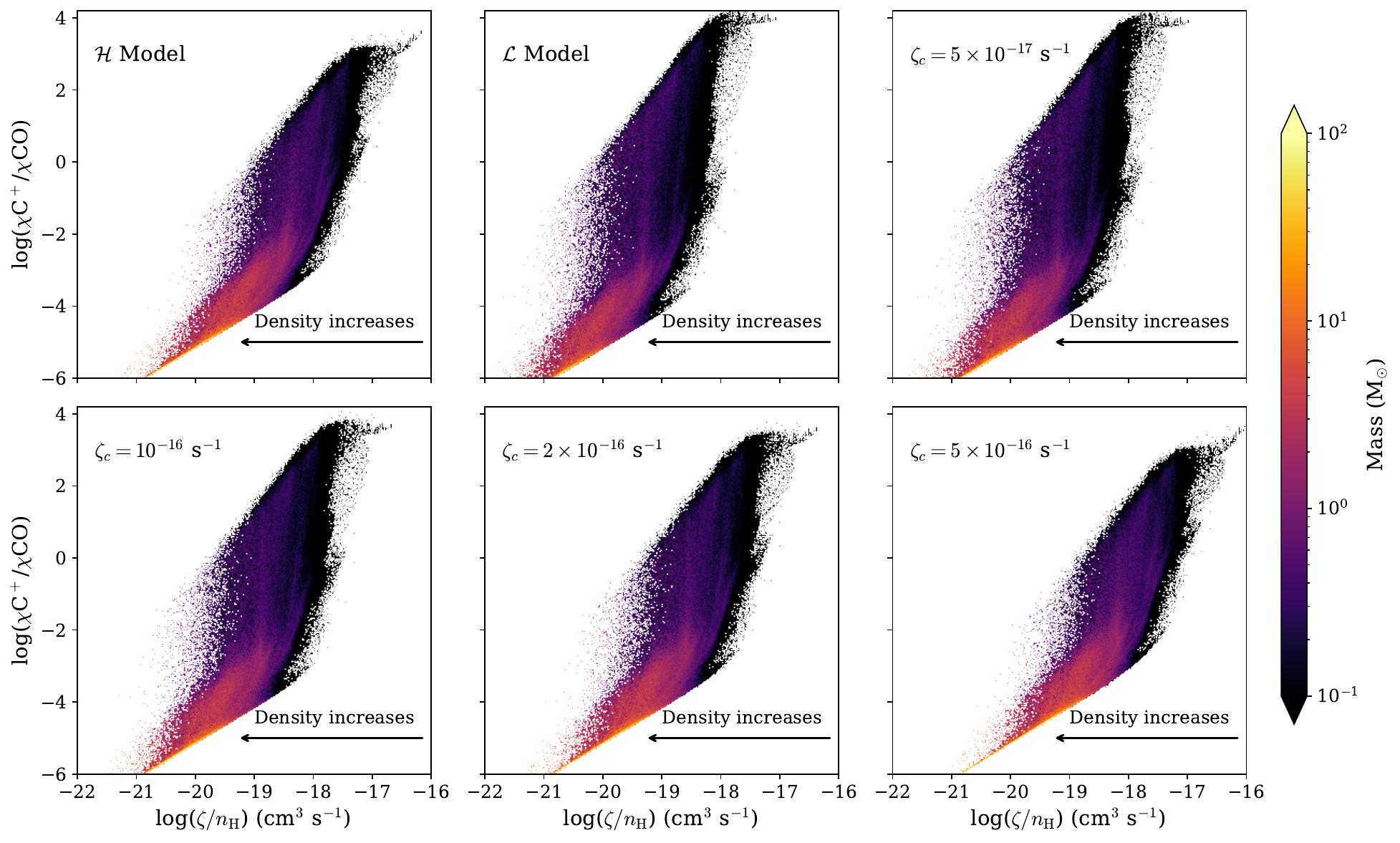}
    \caption{\ce{C+} and \ce{CO} abundance ratio, $\chi(\ce{C+})$/$\chi(\text{CO})$ as a function of $\zeta/n_\text{H}$. Density increases from right to left. The CR model is annotated in each panel. The data is filtered under the constraint: UV $<$ 0.1 $G_0$ and mass-weighted.}
    \label{fig:C+/CO}
\end{figure*}

\begin{figure*}
    \centering
    \includegraphics[width=0.85\linewidth]{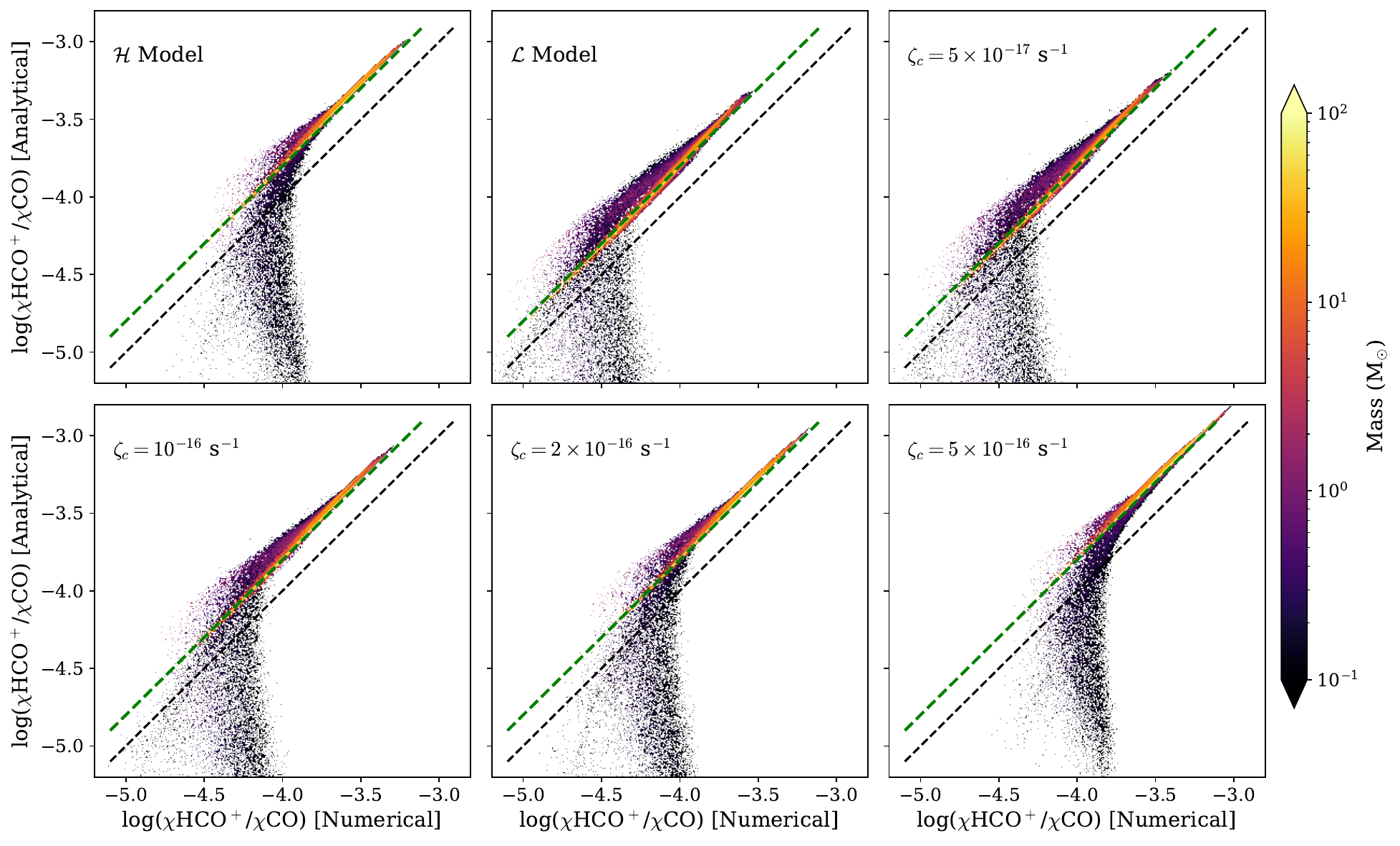}
       \caption{Mass-weighted 1:1 comparison plots between analytical calibrators and numerical results for the abundance ratio, $\chi$(\ce{HCO+})/$\chi$(\ce{CO}). The CR model is annotated in each panel. The black dashed line represents the 1:1 reference, and the green dashed line shows the linearity of the data and the relative deviation in linearity among the models. The data is filtered under the constraints: UV $<$ 0.1 $G_0$ and $n_\text{H} > 10^3$ cm$^{-3}$ for clarity.}
    \label{fig:HCO+/CO_1:1}
\end{figure*}

\begin{figure*}
    \centering
    \includegraphics[width=0.85\linewidth]{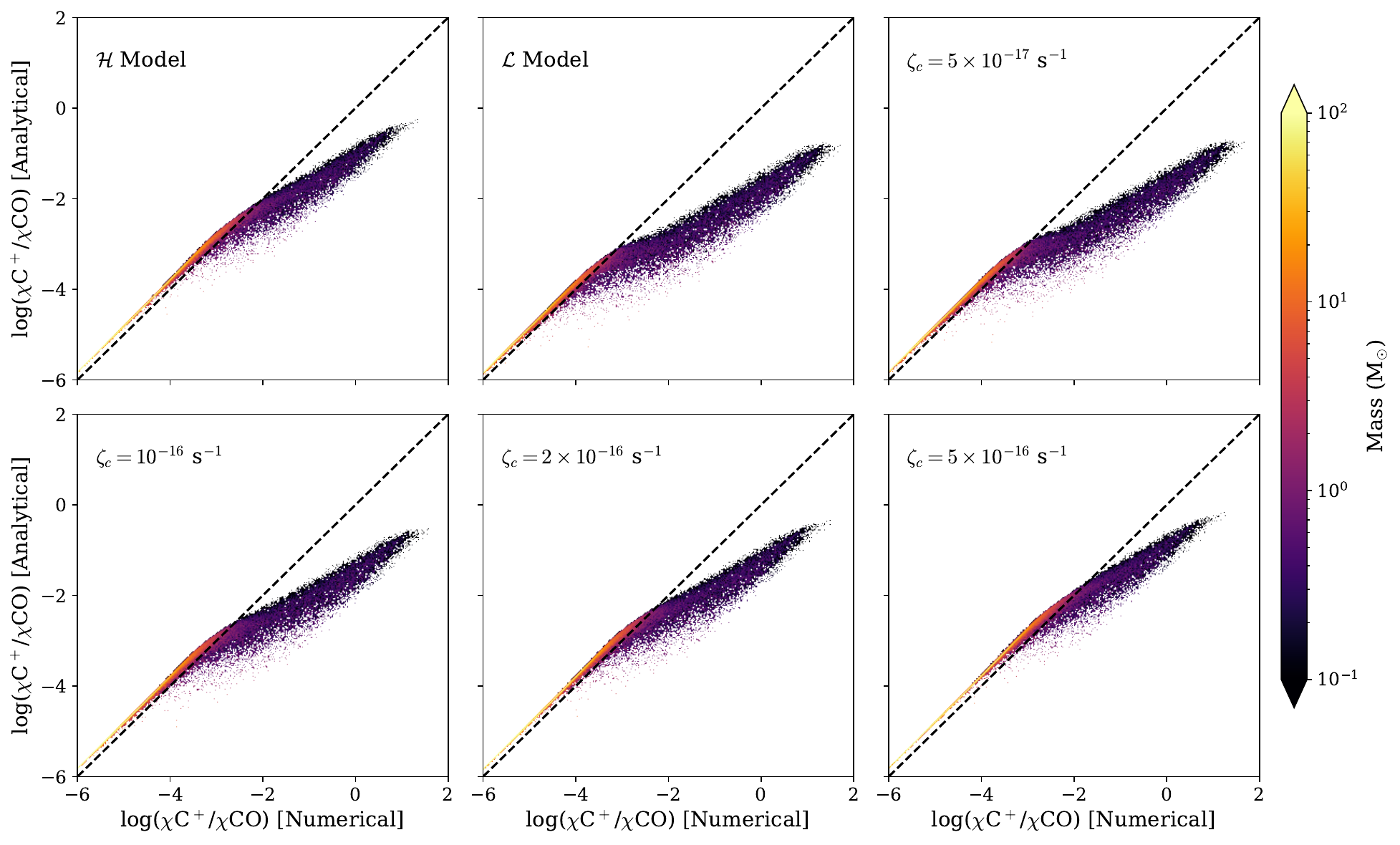}
   \caption{As in Fig. \ref{fig:HCO+/CO_1:1} but for $\chi(\ce{C+})$/$\chi(\text{CO})$. The data is filtered under the constraints: UV $<$ 0.1 $G_0$ and $n_\text{H} > 10^3$ cm$^{-3}$ for clarity.}
    \label{fig:C+/CO_1:1}
\end{figure*}

\begin{figure*}
    \centering
    \includegraphics[width=0.85\linewidth]{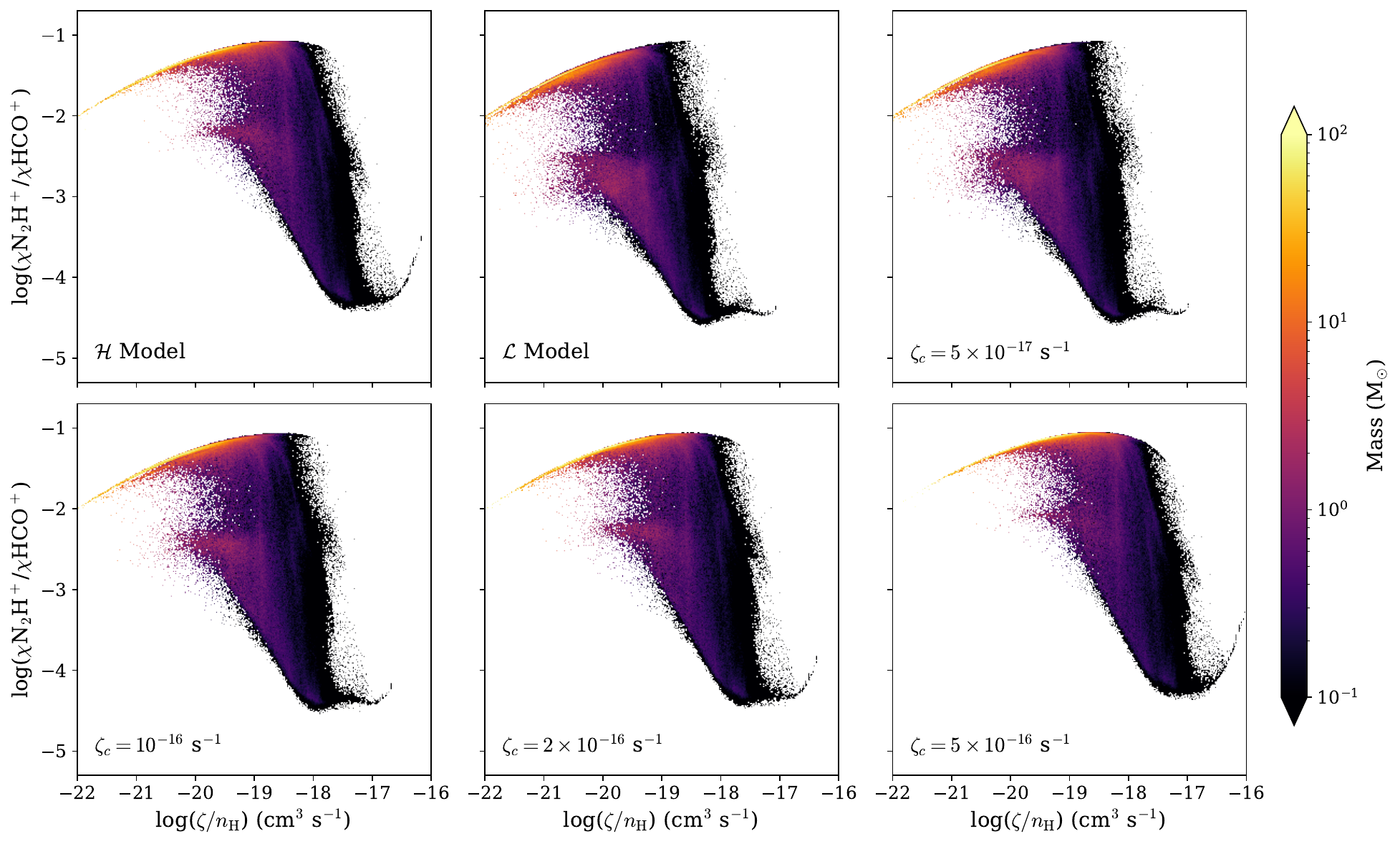}
    \caption{As in Fig. \ref{fig:C+/CO} but for $\chi(\ce{N2H+})$/$\chi(\ce{HCO+})$. Density increases from right to left. The CR model is annotated in each panel. The data is filtered under the constraint: UV $<$ 0.1 $G_0$ and mass-weighted.}
    \label{fig:N2H+/HCO+}
\end{figure*}

The corresponding rate coefficients are listed in Table \ref{tab:rate_coeff}, where $k_e=k_5+k_6$ and $k_\text{CO}=k_7+k_8$. Eq. \ref{eq:1st_order} uses the simple electron-recombination balance and is denoted as \enquote{first-order calibrator}. Eq. \ref{eq:2nd_order} employs a more complex approach to the creation and annihilation of H$_3^+$, utilising different recombination chemistry, and is denoted as \enquote{second-order calibrator}.

The viability of H$_3^+$ calibrators, derived from a first-order expression, is assessed through a comparative analysis of the relative error in H$_3^+$ abundance between analytical and numerical results. Fig. \ref{fig:H3+_relerr} illustrates the error graphs. The first-order calibrator shows very limited performance in capturing the chemistry. In the intermediate regime, the relative error remains small, usually $\leq 0.5$, as most of the data points reside there. But in regions of both low and high density, the relative error becomes significantly larger. Especially at $\zeta/n_\text{H} \leq 10^{-18}\, \text{cm}^3 \,\,\text{s}^{-1}$, the data is very much scattered. In these regions, the relative error can exceed $1.0$, indicating that the first-order calibration struggles to capture the complexity of the chemical interactions. A constant offset of 13\% (marked by a black dashed line) is visible in all the models. The limitations of the first-order expression can, in principle, be minimised by including additional reactions, which become relevant at higher ionization rates, but are not accounted for in this simplified version, leading to substantial deviations between the analytical and numerical solutions. Notably, this comparison is performed with data filtered for UV fields below 1 $G_0$, minimizing the influence of UV radiation.

In contrast, the \enquote{second-order calibrator} incorporates key reactions involving CO and O, which are known to influence H$_3^+$ abundance in environments with higher ionization rates. One notable exclusion in Eq. \ref{eq:2nd_order} is the reaction between \ce{H3+} and carbon (C) atoms, which plays a significant role in the destruction of \ce{H3+}. Neglecting this interaction can lead to discrepancies between analytical and numerical solutions. To address this limitation, we modified the \enquote{second-order calibrator} by incorporating a more comprehensive chemical network, taking into account total oxygen and carbon chemistry. Specifically, we have included \hyperref[tab:rate_coeff]{R10} and \hyperref[tab:rate_coeff]{R11} from Table \ref{tab:rate_coeff}, which account for the interaction between \ce{H3+}, carbon, and oxygen species. This leads to a new coupled analytical expression,
\begin{equation*} 
    \bigg(\frac{\zeta}{n_\text{H}}\bigg)\, \chi(\text{H}_2) = \chi(\text{H}_2^+) [ k_3\chi_e  + k_4\chi (\text{H})  + k_2\chi (\text{H}_2) ],
\end{equation*}
\begin{equation*}  \label{eq:2nd_order_mod} \tag{5}
    k_2\chi (\text{H}_2) \chi(\text{H}_2^+) = \chi(\text{H}_3^+) [k_e\chi_e  + k_{\text{CO}}\chi (\text{CO})  + k_\text{O}\chi (\text{O}) +k_{11}\chi(\text{C})] \,,
\end{equation*}
where $k_{\text{O}}=k_9+k_{10}$. This revised expression performs remarkably well, especially from low to intermediate-dense regions, bringing it close to zero for all the $\zeta$ models, as shown in Fig. \ref{fig:H3+_relerr_mod}. The \enquote{modified second-order calibrator} reduces the relative error to below $30\%$. Here too, a constant shift, now by 6.5\% is observed throughout the models. Notably, the dense region still remains scattered for higher CRIR models, but the error range has been significantly reduced. By integrating a more detailed reaction network, we offer a more accurate and reliable representation of \ce{H3+} chemistry across a broader range of ionization rates.

\subsection{CRIR calibrators using carbon-cycle species}
The $\chi$(\ce{HCO+})/$\chi($\ce{CO}) ratio is a reliable indicator of ionization-driven chemistry. In dense gas, \ce{HCO+} forms primarily through the reaction of \ce{H3+} (produced via ionization of \ce{H2} by CRs; \ref{R1} $-$ \ref{R2}) with \ce{CO} (\hyperref[tab:rate_coeff]{R7}). This reaction directly links \ce{HCO+} formation to the ionization rate. Because \ce{HCO+} is produced through \ce{H3+} interactions with \ce{CO}, in gas where \ce{CO} is more abundant, the increased collision rate enhances \ce{HCO+} production. Thus, without substantial attenuation of the ionization rate, the  $\chi$(\ce{HCO+})/$\chi($\ce{CO}) ratio rises with increasing density.

The $\chi$(\ce{C+})/$\chi$(\ce{CO}) ratio is also sensitive to ionization but reflects the balance between carbon ionization and molecular conversion processes. CRs ionize atomic carbon and other neutral species to form \ce{C+}. At higher $\zeta$, the ionization of neutral carbon is more pronounced alongside the \ce{CO} destruction by \ce{He+} (\hyperref[tab:rate_coeff]{R13}), leading to an increased abundance of \ce{C+} relative to \ce{CO}. 
In dense and shielded regions where CR attenuation becomes significant, the abundance of \ce{C+} decreases as carbon is predominantly locked in \ce{CO}. This makes the $\chi$(\ce{C+})/$\chi$(\ce{CO}) ratio a useful tracer for identifying ionization effects, particularly in regions with varying CR penetration. This relationship, however, is complicated by the fact that \ce{C+} will also be abundant in the envelope of clouds due to UV-driven chemistry, which can influence the overall $\chi$(\ce{C+})/$\chi$(\ce{CO}) ratio.

Fig. \ref{fig:HCO+/CO} and \ref{fig:C+/CO} show $\chi$(\ce{HCO+})/$\chi$(\ce{CO})  as a function of $\zeta/n_e$ and $\chi$(\ce{C+})/$\chi$(\ce{CO})  as a function of $\zeta/n_\text{H}$, respectively. The analysis focuses on data filtered under the constraint that the UV radiation field is below 0.1 $G_0$. This criterion is applied to isolate dense regions where CRs dominate the chemistry, minimizing the effects of UV-driven photochemistry, which prevails in diffuse, low-density environments. To provide a more comprehensive view, the abundance ratios are mass-weighted, emphasizing the dominant processes affecting the bulk of the gas mass rather than the gas volume. This approach accounts for the fact that dense gas contributes significantly to the total mass of molecular clouds but occupies a smaller fraction of the clouds' volume. 

The $\chi$(\ce{HCO+})/$\chi($\ce{CO}) ratio in dense regions exhibits a near-linear behaviour in log–log space, as shown in Fig.~\ref{fig:HCO+/CO} with the pink slope. Most of the gas mass resides in this dense regime, which dominates the trend. In the top-left portion of the phase-space, corresponding to more diffuse regions, the ratio is more scattered and not deterministic, with the total gas mass $< 1$ M$_\odot$. In contrast, the $\chi$(\ce{C+})/$\chi($\ce{CO}) ratio shows a much larger spread across the entire phase-space. Nevertheless, most of the mass again lies in the dense regime. Between $\zeta/n_\mathrm{H} \sim 10^{-20}$ and $10^{-18}$, the ratio undergoes a rapid increase as carbon in the diffuse gas is efficiently ionized to C$^+$, explaining the sharp transition over this range. This behaviour is consistent with the C$^+$ abundance profile as a function of $n_\mathrm{H}$ depicted in Figure~7 of \citet{Gaches2022a}.

Finally, to understand the observed profile trends, we derived analytical expressions for the $\chi$(\ce{HCO+})/$\chi$(\ce{CO}) and $\chi$(\ce{C+})/$\chi$(\ce{CO}), based on the probable dominant steady-state formation and destruction pathways of the respective species. These analytical expressions provide a quantitative framework to interpret the trends, linking the observed ratios to the underlying chemical processes in dense molecular gas. The derived expressions are as follows:
\begin{equation} \tag{6}\label{eq:HCOp/CO}
        \frac{\chi (\text{HCO}^+)} { \chi (\text{CO})} = \frac{k_7k_2[\chi(\text{H}_2)]^2}{k_{12}[k_3\,\chi_e+k_2\,\chi(\text{H}_2)][k_e\,\chi_e + k_7\,\chi(\text{CO})]} \cdot \bigg(\frac{\zeta}{n_e}\bigg)
\end{equation}
\begin{equation} \tag{7}\label{eq:Cp/CO}
        \frac{\chi (\text{C}^+)}{\chi (\text{CO})} = \frac {\mathcal{C}\,k_{13}\,\chi(\text{He})}{[k_{13}\,\chi(\text{CO})+k_{14}\,\chi(\text{O}_2)][k_{15+gr}\,\chi_e+k_{\text{C}^+}\,\chi(\text{O}_2)]} \cdot \bigg(\frac{\zeta}{n_\text{H}}\bigg)\,,
\end{equation}
where, 
\begin{equation} \tag{8}\label{eq:zeta_eff}
\mathcal{C} = \alpha_1' + \alpha_2' ; \quad 
\begin{cases}
\alpha_1' = 0.5\\
\alpha_2' = 0.2/(1.0-0.42) = 0.345
\end{cases}
\end{equation}

\begin{figure}
    \centering
    \includegraphics[width=1.0\linewidth]{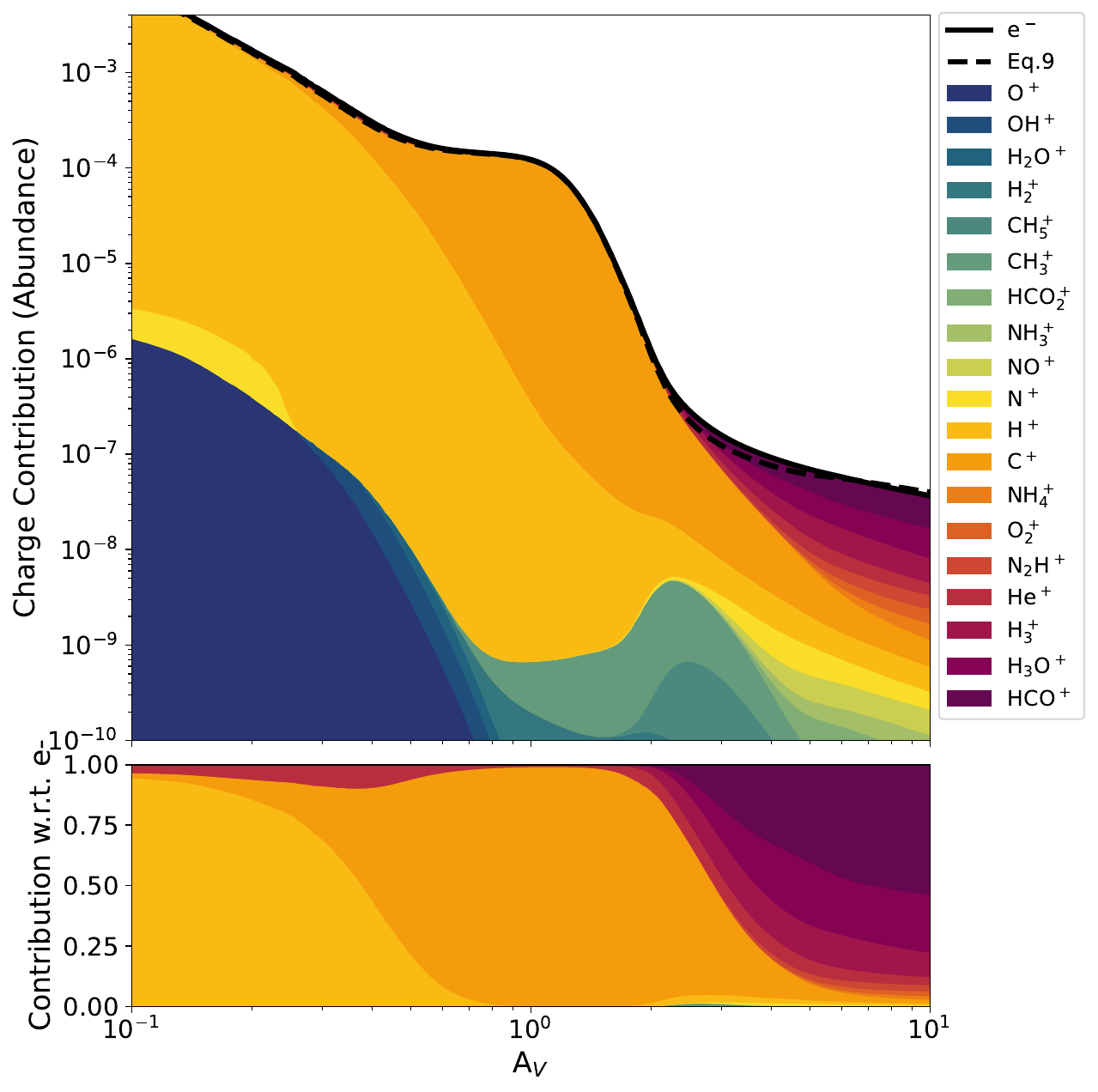}
    \caption{1D $A_V - n_{\rm H}$ slab for charge contribution. The top panel displays the abundance of electrons (solid line) and the corresponding analytical expression (dashed line). The colors show cumulative contributions to the ion balance for positive ions, and the bottom panel shows the normalized cumulative contributions with respect to electron abundance.}
    \label{fig:Av-nH}
\end{figure}

\begin{figure*}
    \centering
    \includegraphics[width=0.85\linewidth]{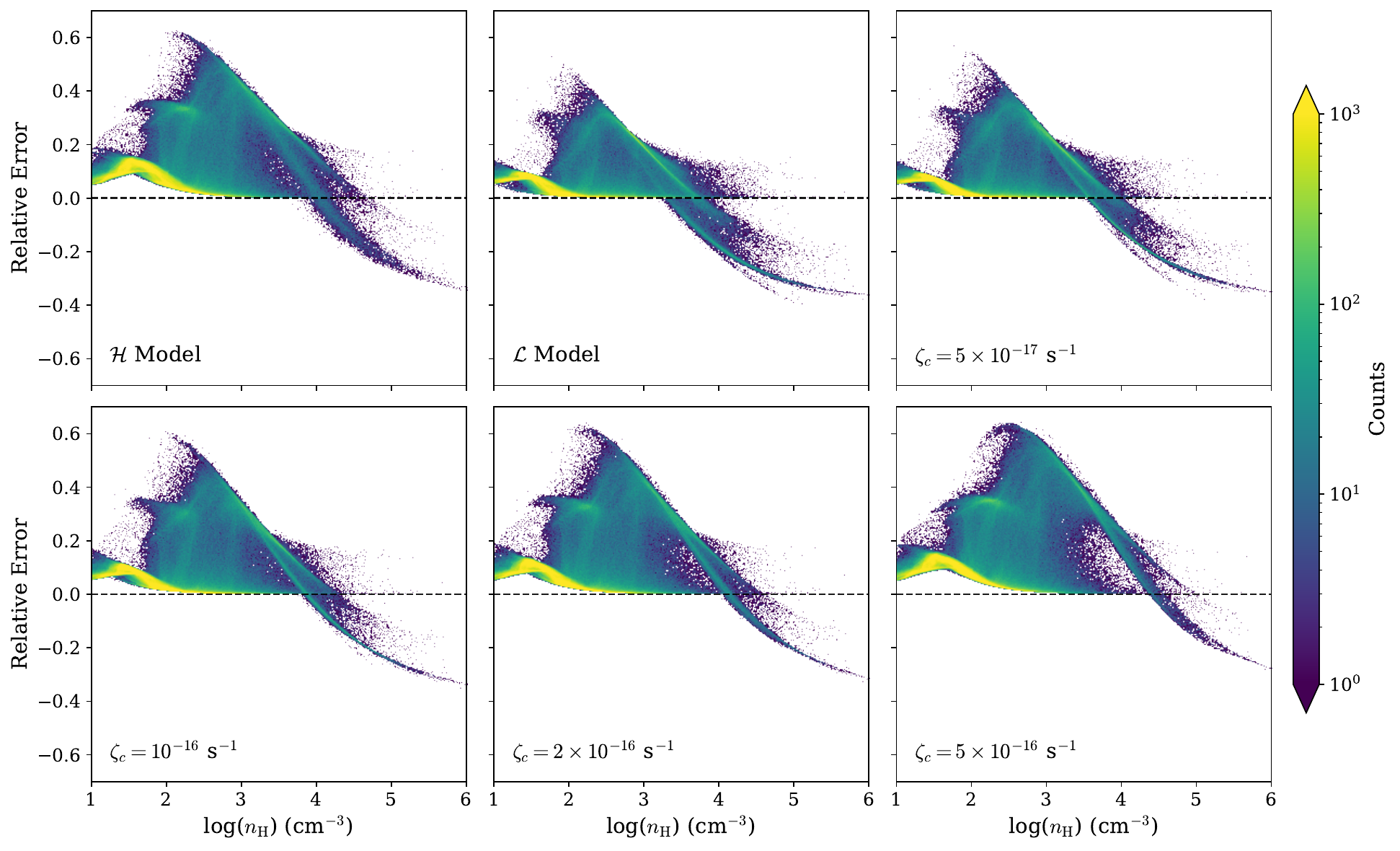}
    \caption{Relative error between the numerical electron fraction ({\sc 3d-pdr}) and the analytical electron fraction expression (Eq.~\ref{eq:xe}).}
    \label{fig:xe_error}
\end{figure*}

The corresponding rate coefficients are listed in Table \ref{tab:rate_coeff}, where $k_e = k_5 + k_6$ and $k_{\text{C}^+} = k_{16} + k_{17}$, $k_{15+gr}$ is the sum of gas phase electron recombination of C$^+$ (\hyperref[tab:rate_coeff]{R15}) and grain assisted recombination, taken from \cite{Gong2017}. The $\alpha_1'$ and $\alpha_2'$ (normalized with respect to $\zeta_0 = 1.3 \times10^{-17}$ s$^{-1}$) comes from the contribution of CRs and CR photons, giving rize to an $\zeta_\text{eff}$ (see Appendix ~\ref{sec:derivation}), for the formation of \ce{He+} from \ce{He}. Figs. \ref{fig:HCO+/CO_1:1} and \ref{fig:C+/CO_1:1} present the mass-weighted 1:1 comparison plots for the analytical calibrator and the numerically filtered data for the abundance ratios of \ce{HCO+}/\ce{CO} and \ce{C+}/\ce{CO}. The black dashed line represents the 1:1 reference line. The $\chi$(\ce{HCO+})/$\chi$(\ce{CO}) plot shows a near-linear trend, suggesting that the derived analytical expression effectively captures the chemistry governing this abundance ratio with a constant offset of 0.2 from the 1.1 line. The 0.2 offset is due to assuming that \ce{H3+} is primarily destroyed by collisions with CO. However, 1D steady-state models showed that depending on the density, the contribution of \ce{H3+ + CO} to the total \ce{H3+} destruction can range between 0.4-0.65, with the rest of the contribution being collisions with O and \ce{N2}, which are difficult to directly observe. Multiplying $k_7$ in the denominator by 2 in Eq.~\ref{eq:HCOp/CO}, or adding reactions between O or \ce{N2} and \ce{H3+} removes this offset. But, both of these species are difficult to directly measure in dense molecular gas. In denser gas, the $k_e \chi_e$ term becomes much less important, so it can be neglected when densities exceed $10^3$ cm$^{-3}$ without significant error.  

In contrast, the $\chi(\ce{C^+})/\chi(\ce{CO})$ plot initially stays parallel with the 1:1 reference line, but towards the right, a slight deviation is observed, indicating that in those regions, the analytical expression fails to predict the abundance ratio compared to the numerical results. Despite this, the majority of the mass remains aligned with the 1:1 line, pointing to a broadly consistent trend throughout the models. The observed deviations are likely due to minor side reactions involving neutral \ce{C} intermediates, which are not integrated in the simplified analytical framework. Notably the approximation of $\chi(\ce{O2})=0.5\times\chi(\ce{CO})\times(\chi(\ce{O})/\chi(\ce{C}))$, where $\chi(\ce{O})$ and $\chi(\ce{C})$ being the standard ISM abundances at solar metallicity, is also a good approximation and plugging it in Eq.~\ref{eq:Cp/CO} gives results depicted in Fig.~\ref{fig:C+/CO_approx}. But we focused with the raw abundance as it is more transparent while comparing the analytical and numerical results. 

As depicted in Fig.~\ref{fig:N2H+/HCO+}, the abundance ratio of \ce{N2H+}/\ce{HCO+} as a function of $\zeta/n_\text{H}$ does not exhibit a clear trend. Interestingly, our results show that the gas-phase production pathways alone can reproduce the \ce{N2H+}/\ce{HCO+} abundance ratio from the modeling without explicitly including a dependence on the CRIR. Both ions are formed and destroyed primarily through a tightly coupled set of reactions---formation via \ce{H3+} collisions, destruction through dissociative recombination with electrons, and the exchange reaction (\hyperref[tab:rate_coeff]{R19}) linking \ce{HCO+} and \ce{N2H+}. These processes operate in a manner such that both species are produced \enquote{in tandem}, and variations in the exchange reaction are largely temperature-driven. Since the temperature itself is only indirectly influenced by $\zeta$, the ratio displays little direct sensitivity to the ionization parameter in our gas-phase--only treatment.

The absence of grain-surface chemistry further contributes to the flatness of the \ce{N2H+}/\ce{HCO+} ratio. In dense regions, freeze-out of species such as CO and subsequent desorption processes---thermal, CR induced, or reactive---strongly alter the abundances of \ce{HCO+} and \ce{N2H+}. These grain-mediated processes break the tight gas-phase coupling and introduce additional pathways that reshape the ratio, particularly by suppressing \ce{HCO+} through CO depletion while enhancing \ce{N2H+}. Such effects have been demonstrated by \citet{Entekhabi2022}, who incorporated CO freeze-out and CR-induced thermal desorption in one-zone infrared dark clouds (IRDC) models and showed that these mechanisms are essential for reproducing observed abundances of \ce{CO}, \ce{HCO+}, and \ce{N2H+}. Including grain chemistry in future work will therefore be crucial for capturing the full behaviour of the \ce{N2H+}/\ce{HCO+} ratio and its true dependence on the ionization environment.

\subsection{Analytical estimation of electron fraction}

A potential caveat in using all the analytical expressions suggested above is the dependence on the electron fraction, $\chi_e$, which cannot be readily measured observationally and must instead be inferred through indirect methods, such as estimates based on recombination rates or ion abundance ratios. A common simplification in astrochemical models is to assume that \ce{HCO+} is a primary carrier of positive charge, often adopting the approximation \(\chi(\ce{HCO+})/\chi_{\ce{e}} \sim 0.5\). While this serves as a practical constraint for CRIR estimation, it may not capture the full complexity of ionization processes across different environments. Although analytical expressions for estimating the $\chi_e$ have been proposed \citep{Feng2016,Luo_2024,Latrille2025}, a robust, general estimator valid across a wide density range remains lacking, mainly due to the inclusion of \ce{N2H+} as a charge carrier in previous approximations.
 
Recent work by \cite{Pineda2024}, which mapped the NGC 1333 SE region, revealed a smooth $\chi_e$ distribution with a typical median value of $10^{-6.5}$. This suggests that the actual $\chi_e$ values are significantly higher than the \ce{HCO+} abundance, highlighting the limitations of the approximation $\chi_e = 2\chi(\ce{HCO+})$. While some studies use the alternative approximation \(\chi_e = \chi(\ce{C+})\), this approach tends to underestimate \(\chi_e\) in regions where the CRIR is high enough for \(\ce{H+}\) to become the dominant source of electrons \citep{Indriolo2012a,Indriolo12b,Indriolo2015,Bialy2015}.

To address these limitations, we propose a more accurate representation of the electron fraction given by,
\begin{equation*} \tag{9} \label{eq:xe} \chi_e = \chi(\ce{H+})+\chi(\ce{C+})+2\chi(\ce{HCO+}) \end{equation*}

This formulation provides a better estimate of $\chi_e$, as shown by the one-dimensional model results in Fig.~\ref{fig:Av-nH}. The one-dimensional model uses an $A_V - n_{\rm H}$ density distribution that reproduces the behaviors seen in three-dimensional simulations, namely exhibiting low (high) density gas at low (high) extinction, see Appendix \ref{network_benchmark} and \citet{Bisbas2019} and \citet{Gaches2025}. Additionally, it remains within a $\pm60\%$ error range in 3D space with $-30\%$ in higher density regions, as depicted in Fig.~\ref{fig:xe_error}. This approach offers a more reliable method for estimating CRIR in astrochemical models, though further refinements are necessary to develop a robust analytical estimator for $\chi_e$.

\section{Conclusions and outlook} \label{conclusion}

Building upon the work of \citep{Gaches2022a}, we have expanded the previous study by incorporating a more comprehensive chemical network that includes nitrogen chemistry. Additionally, we have investigated the impact of CR attenuation, $\zeta(N)$, through both the $\mathcal{H}$ and $\mathcal{L}$ models. Our key findings are as follows:
\begin{enumerate}
 \item The numerical results successfully reproduce the steady-state analytic \ce{H3+} abundances with an error below $30\%$, consistent with the hydrogen-electron balance analytical approach used to constrain the CRIR. The revised expression performs remarkably well, particularly from low to intermediate-density regions, bringing the error close to zero (see Fig. \ref{fig:H3+_relerr_mod}). By incorporating a more detailed reaction network, this approach provides a more accurate and reliable representation of \ce{H3+} chemistry across a broader range of ionization conditions.

 \item    The abundance ratios \ce{HCO+}/CO and \ce{C+}/CO exhibit a highly correlated relationship across all CR models, particularly in dense molecular regions, making them robust tracers of ionization conditions. The numerical results closely align with analytical calibrations, although the deviation from the 1:1 line in \ce{C+}/CO (see Fig. \ref{fig:C+/CO_1:1}) suggests additional side reactions not fully captured by the simplified chemistry.

 \item    In contrast, the \ce{N2H+}/\ce{HCO+} abundance ratio does not exhibit a clear trend as a function of $\zeta/n_\text{H}$, likely due to the absence of grain-chemistry in our model. This highlights the significance of surface reactions in determining ion abundances, particularly in the densest cloud environments ($n_\text{H} > 10^4$ cm$^{-3}$).

 \item    The revised formulation for estimating the $\chi_e$ provides an improved estimate compared to an $A_V-n_\text{H}$ PDR model and maintains an error margin within $\pm 60\%$ in 3D phase-space. This suggests it may be a more reliable prescription for CRIR in astrochemical models, though further refinements are required to develop a robust analytical estimator.

\end{enumerate}
The low CRIR values inferred in IRDCs by \citet{Entekhabi2022} may themselves be a signature of attenuation processes, further motivating the kind of physically driven modeling approach presented in this work. Future studies need to include grain-surface processes such as CO freeze-out and CR-induced desorption to better reflect the physical conditions of dense regions such as IRDCs, as also emphasized by \citet{Entekhabi2022}. These would further enhance the model’s ability to reproduce observed molecular trends in IRDCs and other dense regions of the interstellar medium.

In summary, our results provide a detailed investigation of CR attenuation effects on molecular ionization chemistry and abundance profiles. The use of a column density-dependent CRIR prescription significantly improves agreement with observational diagnostics and highlights the limitations of constant ionization rate models. Our findings support the need for non-uniform CRIR treatments in dense molecular clouds.

\begin{acknowledgements}
The authors thank the anonymous referee for the comments, which improved the clarity of this work. AR acknowledges support from the Chalmers Astrophysics and Space Sciences Summer (CASSUM) research fellowship. BALG is supported by the German Research Foundation (DFG) in the form of an Emmy Noether Research Group - DFG project \#542802847 (GA 3170/3-1). JCT acknowledges support from ERC Advanced grant 788828 (MSTAR).  The computations were performed on the Dardel/PDC supercomputing facility, enabled by resources provided by the National Academic Infrastructure for Supercomputing in Sweden (NAISS), partially funded by the Swedish Research Council through grant agreement no. 2022-06725 through the allocations NAISS 2024/1-27 and NAISS 2024/6-292. The authors gratefully acknowledge the computing time granted by the Center for Computational Sciences and Simulation (CCSS) of the University of Duisburg-Essen and provided on the supercomputer amplitUDE (DFG project 459398823; grant ID INST 20876/423-1 FUGG) at the Center for Information and Media Services (ZIM).
\end{acknowledgements}
%
%-------------------------------------------------------------------

\bibliographystyle{aa}
\bibliography{references}

\begin{appendix}
\onecolumn

\section{Abundance profiles and relative errors}
\begin{figure}[H]
    \centering
    \includegraphics[width=0.85\linewidth]{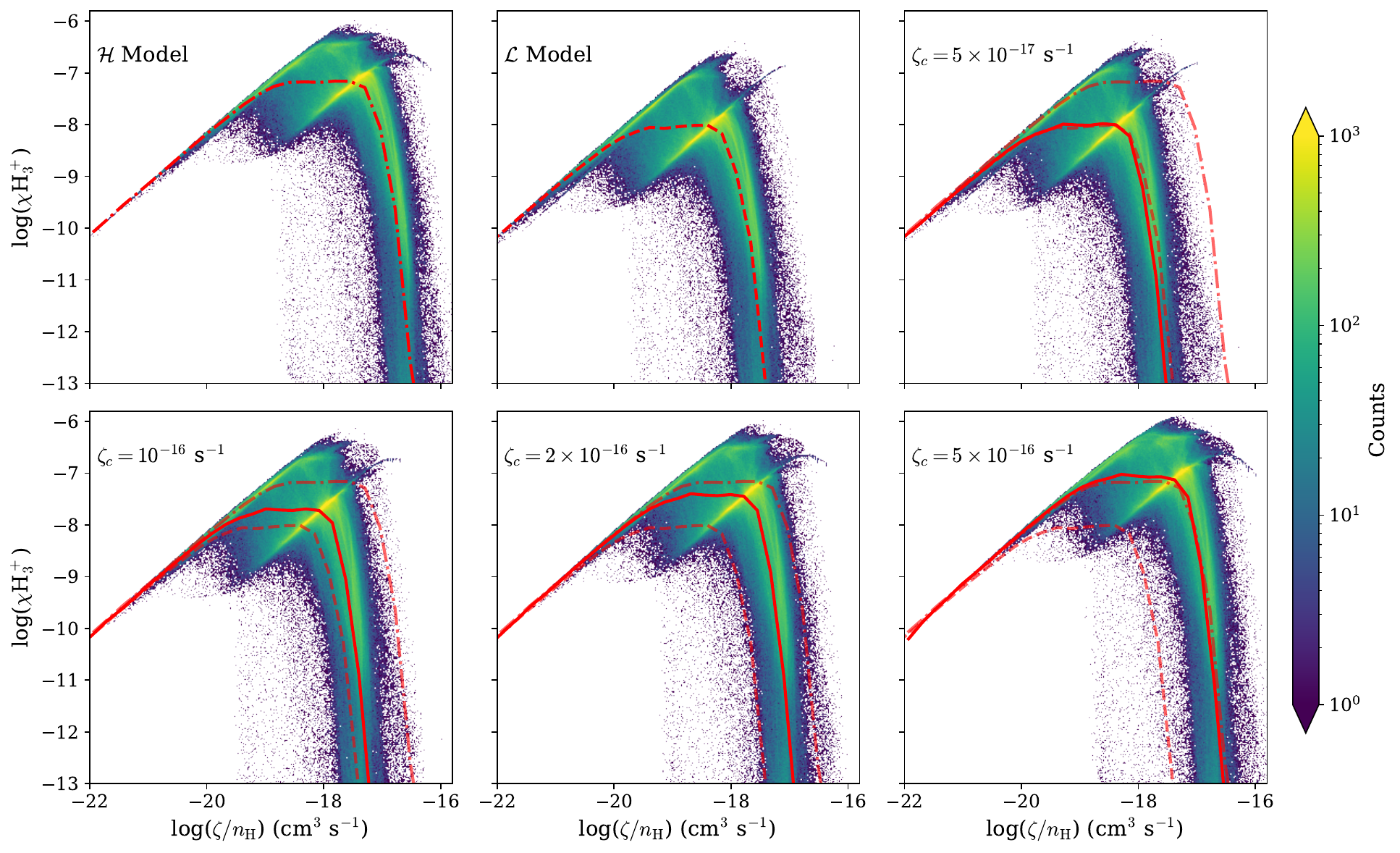}
    \caption{As in Fig. \ref{fig:CO_abund}, but for \ce{H3+} abundance profile. Density increases from right to left.}
    \label{fig:H3+_abundace}
\end{figure}

\begin{figure}[H]
    \centering
    \includegraphics[width=0.85\linewidth]{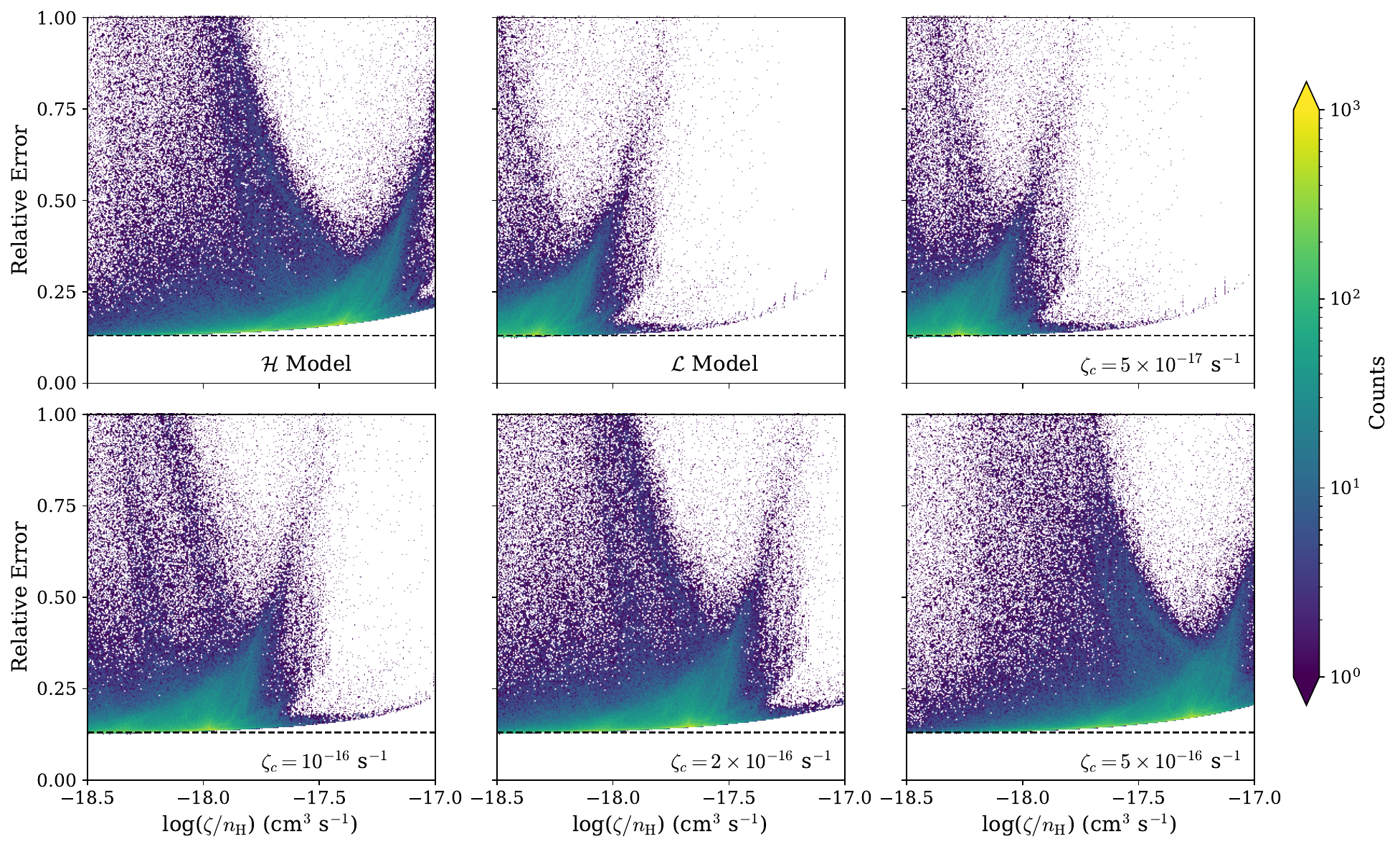}
    \caption{Relative error between the numerical abundance profile ({\sc 3d-pdr}) and first-order analytical expression of H$_3^+$ abundance, calculated using Eq. \ref{eq:1st_order} from \cite{Indriolo2012a}. The constant offset of 13\% is denoted by the black dashed line.}
    \label{fig:H3+_relerr}
\end{figure}

\begin{figure}
    \centering
    \includegraphics[width=0.85\linewidth]{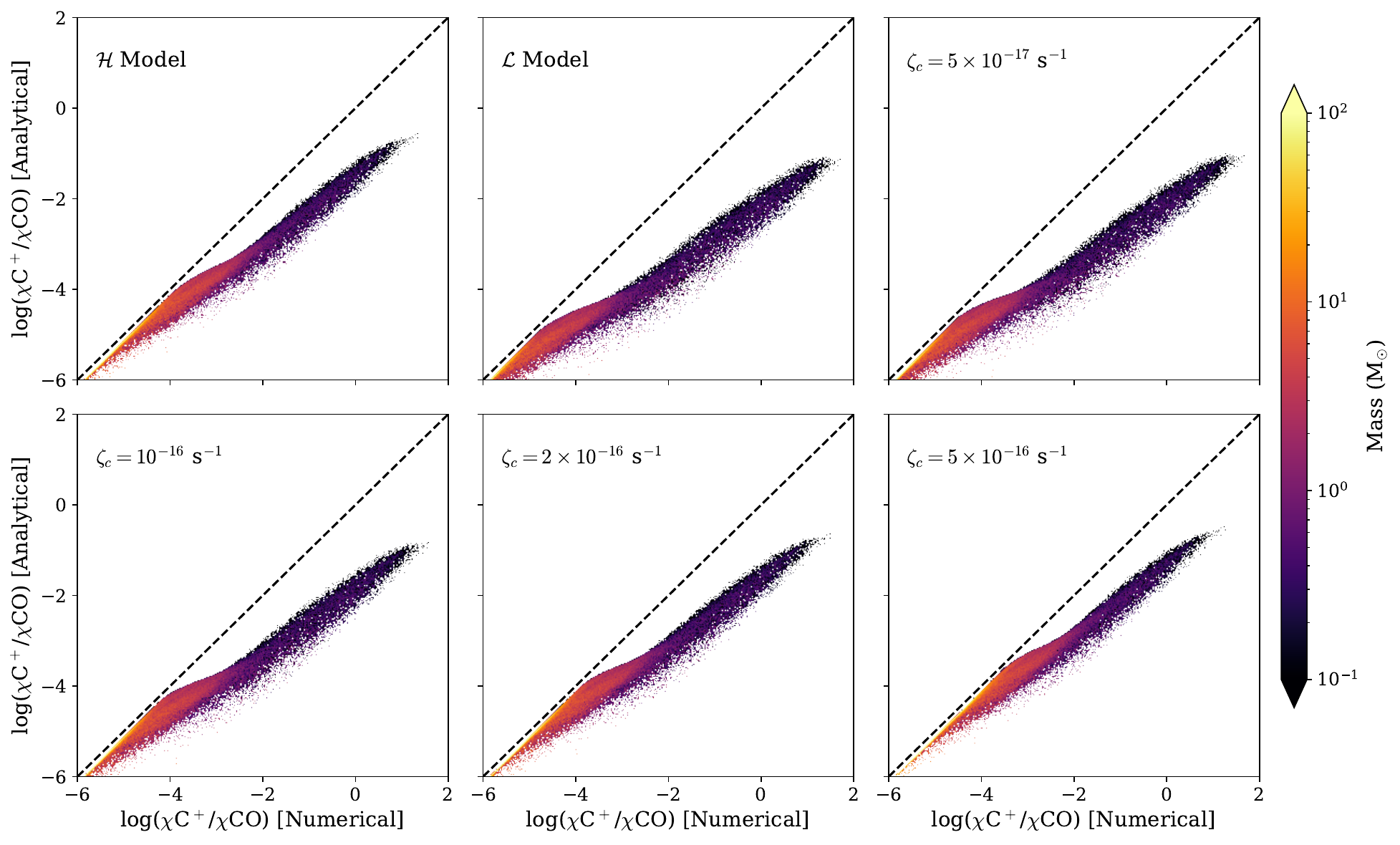}
    \caption{As in Fig. \ref{fig:C+/CO_1:1} but for the approximate abundance of $\chi(\ce{O2})$. The data is filtered under the constraints: UV $<$ 0.1 $G_0$ and $n_\text{H} > 10^3$ cm$^{-3}$ for clarity.}
    \label{fig:C+/CO_approx}
\end{figure}

\begin{figure}
    \centering
    \includegraphics[width=1.0\linewidth]{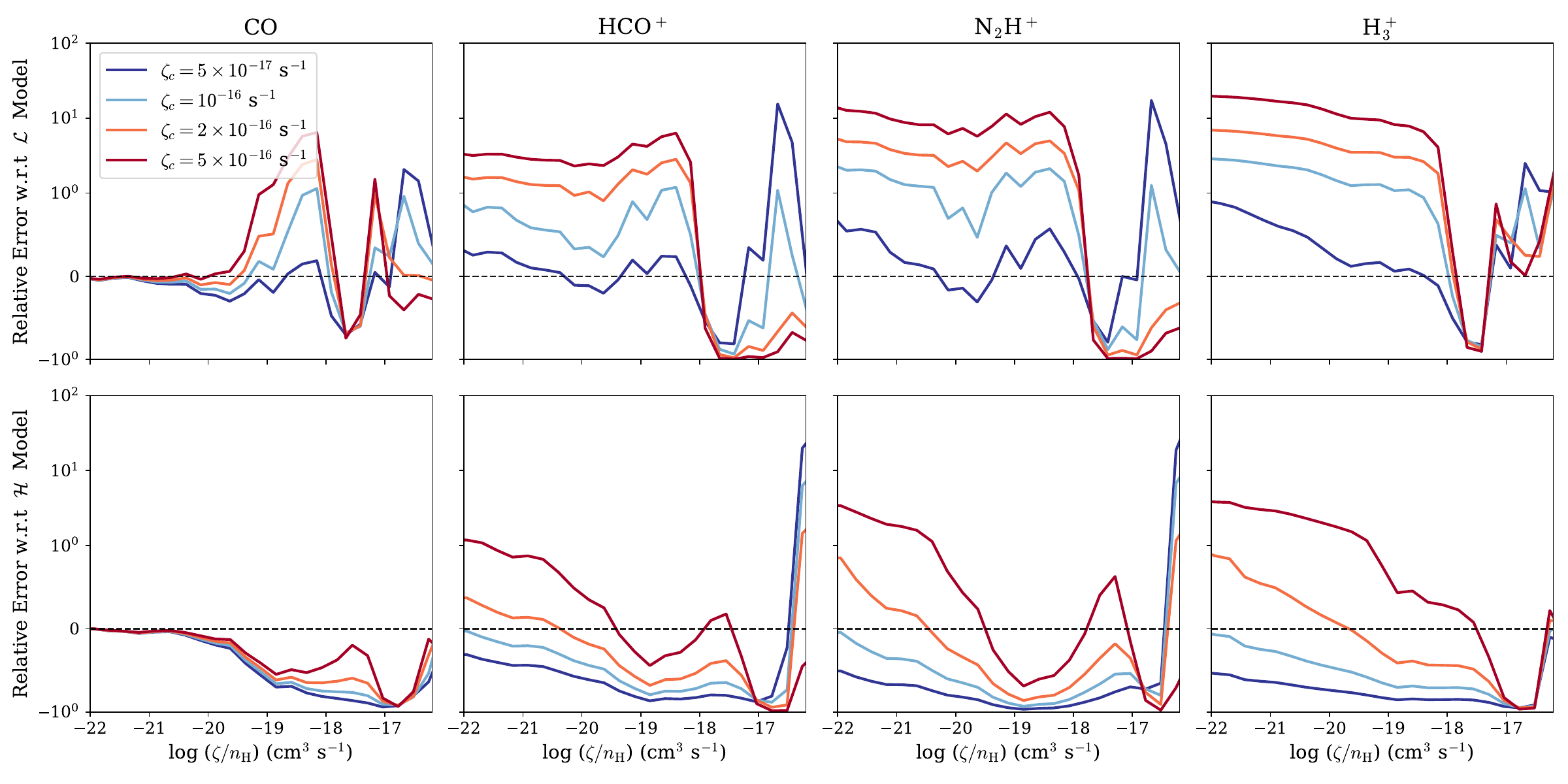}
    \caption{ The linear-relative error in the average abundance profiles for a particular constant CRIR model in comparison to the $\zeta(N)_\mathcal{L}$ (\textit{top}) and $\zeta(N)_\mathcal{H}$ (\textit{bottom}) model, calculated using Eq. \ref{eq:rel_Err}.}
    \label{fig:all_relerr}
\end{figure}

The following figures provide an overview of the key abundance profiles and relative errors discussed in the main text. Fig.~\ref{fig:H3+_abundace} shows the H$_3^+$ abundance profile, $\chi(\mathrm{H}_3^+)$, versus the effective ionization rate, $\zeta / n_\mathrm{H}$, density increasing from right to left. Fig.~\ref{fig:H3+_relerr} presents the relative error between the numerical {\sc 3d-pdr} abundance profile and the first-order analytical expression of H$_3^+$ \citep[Eq.~\ref{eq:1st_order} from][]{Indriolo2012a}. Fig.~\ref{fig:C+/CO_approx} shows the approximate $\chi(\mathrm{C^+}) / \chi(\mathrm{CO})$ 1:1 phase plot. Fig.~\ref{fig:all_relerr} displays the linear-relative error in the average abundance profiles for constant CRIR models compared to $\zeta(N)_\mathcal{L}$ (top) and $\zeta(N)_\mathcal{H}$ (bottom).

\clearpage

\section{Derivation of the carbon-cycle calibrators}
\subsection{Steady-state derivation of the HCO$^+$/CO abundance ratio}

We consider the following key ion--molecule reactions in the ISM (reaction and rates no. are same as Table~\ref{tab:rate_coeff}):
\begin{align}
&\mathrm{H_2} + \mathrm{CR} \rightarrow \mathrm{H_2^+} + e^- \quad (\zeta) \\
&\mathrm{H_2^+} + \mathrm{H_2} \rightarrow \mathrm{H_3^+} + \mathrm{H} \quad (k_2) \\
&\mathrm{H_2^+} + e^- \rightarrow \mathrm{H} + \mathrm{H} \quad (k_3) \\
&\mathrm{H_3^+} + e^- \rightarrow \text{products} \quad (k_5+k_6=k_e) \\
&\mathrm{H_3^+} + \mathrm{CO} \rightarrow \mathrm{HCO^+} + \mathrm{H_2} \quad (k_7) \\
&\mathrm{HCO^+} + e^- \rightarrow \mathrm{CO} + \mathrm{H} \quad (k_{12})
\end{align}
Under steady-state conditions for each ion:
\paragraph{($i$) For $\mathrm{H_2^+}$:}
\begin{equation}
\zeta\, n(\mathrm{H_2}) = n(\mathrm{H_2^+}) \left[ k_2\,n(\mathrm{H_2}) + k_3\,n_e \right] \implies n(\mathrm{H_2^+}) = \frac{\zeta\, n(\mathrm{H_2})}{k_2\,n(\mathrm{H_2}) + k_3\,n_e}
\end{equation}
\paragraph{($ii$) For $\mathrm{H_3^+}$:}
Formation via $\mathrm{H_2^+ + H_2}$ and destruction via CO and $e^-$ recombination yield
\begin{equation}
k_2\, n(\mathrm{H_2^+})\, n(\mathrm{H_2}) = n(\mathrm{H_3^+})\left[k_7\,n(\mathrm{CO}) + k_e\,n_e\right]
\end{equation}
Substituting $n(\mathrm{H_2^+})$ from above gives
\begin{equation} \label{eq:b09}
n(\mathrm{H_3^+}) = 
\frac{k_2\, \zeta\, [n(\mathrm{H_2})]^2}{
\left[k_2\,n(\mathrm{H_2}) + k_3\,n_e\right]
\left[k_7\,n(\mathrm{CO}) + k_e\,n_e\right]}
\end{equation}
\paragraph{($iii$) For $\mathrm{HCO^+}$:}
Formation via $\mathrm{H_3^+ + CO}$ and destruction via dissociative recombination lead to
\begin{equation}
k_7\,n(\mathrm{H_3^+})\,n(\mathrm{CO}) = k_{12}\,n(\mathrm{HCO^+})\,n_e \implies \frac{n(\mathrm{HCO^+})}{n(\mathrm{CO})}
= \frac{k_7\,n(\mathrm{H_3^+})}{k_{12}\,n_e}
\end{equation}
Substituting $n(\mathrm{H_3^+})$ from Eq.~(\ref{eq:b09}) gives
\begin{equation}
\frac{n(\mathrm{HCO^+})}{n(\mathrm{CO})}
= \frac{k_7\,k_2\,\zeta\,[n(\mathrm{H_2})]^2}{
k_{12}\,n_e\,
\left[k_2\,n(\mathrm{H_2}) + k_3\,n_e\right]
\left[k_7\,n(\mathrm{CO}) + k_e\,n_e\right]}
\end{equation}
\paragraph{($iv$) In terms of fractional abundances:}
Defining $\chi(i)=n(i)/n_\text{H}$ and $\chi_e=n_e/n_\text{H}$, we obtain
\begin{equation}
\boxed{
\frac{\chi(\mathrm{HCO^+})}{\chi(\mathrm{CO})}
= 
\frac{k_7\,k_2\,[\chi(\mathrm{H_2})]^2}{
k_{12}\,
\left[k_2\,\chi(\mathrm{H_2}) + k_3\,\chi_e\right]
\left[k_7\,\chi(\mathrm{CO}) + k_e\,\chi_e\right]}
\cdot\left(\frac{\zeta}{n_e}\right)
}
\end{equation}

\subsection{Steady-state derivation of the C$^+$/CO abundance ratio}  \label{sec:derivation}
We consider the following key reactions for Helium ionization:
\begin{align}
& \mathrm{He} + \mathrm{CRP} \;\rightarrow\; \mathrm{He^+} + e^- 
\quad (\zeta_{\mathrm{CRP}}=\alpha_1\,\zeta) \\
& \mathrm{He} + \mathrm{CRPHOT} \;\rightarrow\; \mathrm{He^+} + e^- 
\quad (\zeta_{\mathrm{CRPHOT}}= \alpha_2\,\zeta)
\end{align}
The ionization of helium is therefore initiated through two distinct pathways: 
(i) direct ionization by cosmic-ray particles (CRP) and 
(ii) ionization by secondary cosmic-ray--induced photons (CRPHOT), 
with rate coefficients 
$\alpha_1 = 6.5\times10^{-18}\,\mathrm{s^{-1}}$ and 
$\alpha_2 = (1.3\times10^{-17}) \times (0.2/[1-0.42]) \;\mathrm{s^{-1}}$, respectively.

To express these in normalized form relative to the reference cosmic-ray ionization rate 
$\zeta_0 = 1.3\times10^{-17}\,\mathrm{s^{-1}}$, 
we define the dimensionless parameters
\begin{equation}
\alpha'_1 = \frac{\alpha_1}{\zeta_0}, 
\qquad
\alpha'_2 = \frac{\alpha_2}{\zeta_0},
\end{equation}
yielding 
$\alpha'_1 = 0.5$ and $\alpha'_2 \approx 0.345$.
The total effective ionization rate of helium is then
\begin{equation}
\boxed{\zeta_{\mathrm{eff}} = \zeta\,(\alpha'_1 + \alpha'_2) = \zeta\,\mathcal{C}}\,,
\end{equation}
where $\zeta$ is the primary cosmic-ray ionization rate of $\mathrm{H_2}$.
\clearpage
Hence, for the calculation of the steady-state abundance ratio of \(\mathrm{C^+}/\mathrm{CO}\), we consider the following key reactions in the gas-phase network:
\begin{align}
& \mathrm{He} + \mathrm{CR} \;\rightarrow\; \mathrm{He^+} + e^- 
\quad (\mathcal{C}\,\zeta) \\
& \mathrm{He^+} + \mathrm{CO} \;\rightarrow\; \mathrm{C^+} + \mathrm{O} + \mathrm{He}
\quad (k_{13}) \\
& \mathrm{He^+} + \mathrm{O_2} \;\rightarrow\; \mathrm{O^+} + \mathrm{O} + \mathrm{He}
\quad (k_{14}) \\
& \mathrm{C^+} + e^- \;\rightarrow\; \mathrm{C} + h\nu 
\quad (k_{15}) \\
& \mathrm{C^+} + \mathrm{O_2} \;\rightarrow\; \text{products}
\quad (k_{\text{C}^+}=k_{16}+k_{17})
\end{align}
Additionally, including the grain-mediated recombination reactions from \cite{Gong2017}, for the reaction $\mathrm{C^+} + e^- + \mathrm{gr} \;\rightarrow\; \mathrm{C} + \mathrm{gr}$, we get :
\[
\alpha_{\mathrm{g}}
= 10^{-14} \,
\frac{C_0}
{1 + C_1\,\Psi^{C_2}
+ C_1\,\Psi^{\,C_2 - C_5 - C_6 \ln T}\, C_3\,T^{C_4}}\,,
\]
where,
\[
\Psi = 
1.68 \, \mathrm{UV_{ext}}\,
\exp\,\!\left[-\,\mathrm{UV_{fac}} \, A_{V,\text{eff}}\right]\,
\frac{\sqrt{T}}{n_e}
\]
and the constants are,
\[
C_0 = 45.58,\,
C_1 = 6.089\times 10^{-3}, \,
C_2 = 1.128, \,
C_3 = 4.331\times 10^{2}, \,
C_4 = 4.845\times 10^{-2}, \,
C_5 = 0.8120, \,
C_6 = 1.333\times 10^{-4},
\]
\[\mathrm{UV_{ext}} = 10 \,G_0,\,
\mathrm{UV_{fac}} = 3.02\]
Hence, the Total rate for electron recombination, $k_{15+gr}=k_{15}+(\alpha_{\mathrm{g}}/\chi_e)$; the division with $\chi_e$ is there as a normalization factor, converting the grain-assisted recombination rate from the formulation used by \cite{Gong2017} into a standard two-body rate coefficient compatible with the rest of our chemical network. \\
\\
Under steady-state conditions for each ion:
\paragraph{($i$) For $\mathrm{He^+}$:}
\begin{equation} \label{eq:b22}
\mathcal{C}\,\zeta\, n(\mathrm{He}) = n(\mathrm{He^+}) \left[ k_{13}\,n(\mathrm{CO}) + k_{14}\,n(\mathrm{O_2}) \right] \implies n(\mathrm{He^+}) = \frac{\mathcal{C}\,\zeta\, n(\mathrm{He})}{ k_{13}\,n(\mathrm{CO}) + k_{14}\,n(\mathrm{O_2})}
\end{equation}
\paragraph{($ii$) For $\mathrm{C^+}$:}
Formation via $\mathrm{He^+ + CO}$ and destruction via gas-phase and grain mediated recombination alongside $\mathrm{C^+ + O_2}$ reactions lead to
\begin{equation}
n(\ce{C+})(k_\mathrm{C^+}\, n(\ce{O2}) + k_{15+gr}\, n_e)
= k_{13}\,n(\ce{He+})\,n(\ce{CO}) \implies  \frac{n(\mathrm{C^+})}{n(\mathrm{CO})} = \frac{k_{13}\,n(\mathrm{He^+})}{k_\mathrm{C^+}\, n(\ce{O2}) + k_{15+gr} \,n_e}
\end{equation}
Substituting $n(\mathrm{He^+})$ from Eq.~(\ref{eq:b22}) gives
\begin{equation}
\frac{n(\mathrm{C^+})}{n(\mathrm{CO})}
= \frac{k_{13}\,\mathcal{C}\,\zeta\,n(\mathrm{He})}{
\left[k_{13}\,n(\mathrm{CO}) + k_{14}\,n(\mathrm{O_2})\right]
\left[k_\mathrm{C^+}\, n(\ce{O2}) + k_{15+gr} \,n_e\right]}
\end{equation}
\paragraph{($iii$) In terms of fractional abundances:}
Defining $\chi(i)=n(i)/n_\text{H}$ and $\chi_e=n_e/n_\text{H}$, we obtain
\begin{equation}
\boxed{
\frac{\chi(\mathrm{C^+})}{\chi(\mathrm{CO})}
= \frac{k_{13}\,\mathcal{C}\,\chi(\mathrm{He})}{
\left[k_{13}\,\chi(\mathrm{CO}) + k_{14}\,\chi(\mathrm{O_2})\right]
\left[k_\mathrm{C^+}\, \chi(\ce{O2}) + k_{15+gr} \,\chi_e\right]}\cdot\left(\frac{\zeta}{n_\text{H}}\right)
}
\end{equation}

\section{Description of the chemical network and steady-state benchmarks} \label{network_benchmark}
We present here a general description of the chemical network that was used in the study and the benchmarks of the network that were performed. The network is a subset of UMIST2012 in size between the reduced networks previously used (33 species) and the full network ($>$200 species), with 77 species primarily with 4 atoms or fewer with a few exceptions for molecular ions necessary for the desired chemistry (\ce{NH4+}, \ce{H3CO+}, \ce{C2H3}, and \ce{CH5+}). The primary goal of the network was to model the gas-phase chemistry of \ce{HCN}, \ce{HNC}, \ce{N2H+}, \ce{NH3}, and light hydrocarbons such as \ce{C2H}. The network was constructed by taking the full-sized network and cutting it down in species and reactions in size until a balance was achieved between performance and accuracy. In practice, this was carried out by removing all species and reactions with molecules with more than four atoms and species bearing atoms beyond H/He/C/N/O. Then, species and reactions were added back in manually as needed for the chemistry. The network was benchmarked against the reduced and full network currently in {\sc 3d-pdr} by computing a grid of 9 different models, consisting of three FUV fields, $G_0 = 1, 10, 100$ in units of the Draine field \citep{Draine1978}, and three cosmic-ray ionization rates, $\zeta = 10^{-17}, 10^{-16}, 10^{-15}$. The benchmark was performed using the ``$A_V - n_{\rm H}$'' density distribution \citep{Bisbas2019, Gaches2025}, which reproduces the relation
\begin{equation}
    A_{V, {\rm eff}} = 0.05 \exp{\left [ 1.6 \left ( \frac{n_{\rm H}}{{\rm cm^{-3}}} \right )^{0.12} \right ]} ~{\rm mag}
\end{equation}
This density distribution has the benefit that it probes realistic regimes with low-density, highly irradiated environments and dense, shielded gas. Since {\sc 3d-pdr} integrates the chemistry to steady state, the benchmarking is performed under this assumption. We emphasize that this network was developed to best reproduce the behavior of the full gas-phase network in average molecular cloud environments, e.g., $n_\text{H} < 10^4$ cm$^{-3}$, and does not include freeze out at this time. The network will be made available with the next public release of {\sc 3d-pdr}, or upon request.

\begin{figure}
    \centering
    \includegraphics[width=\textwidth]{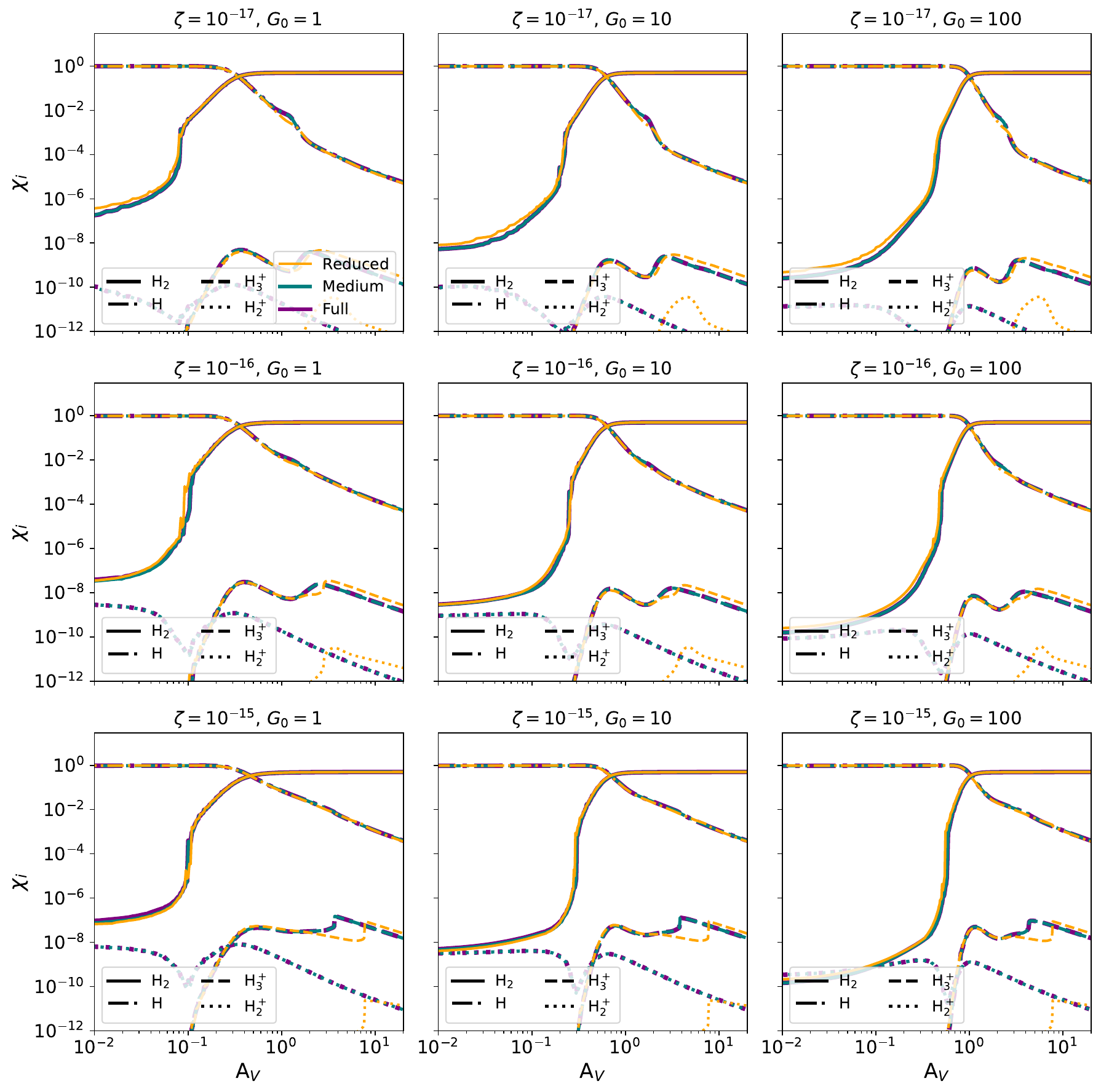}
    \caption{\label{fig:HmedBench}The abundance of the denoted species, $\chi_i$, versus visual extinction, $A_V$. Benchmark of the medium-sized chemical network (teal) against the reduced network (orange) and full network (purple), focusing on molecular hydrogen chemistry. The external FUV radiation field, in units of the Draine field \citep{Draine1978}, and cosmic-ray ionization rate are annotated in the subfigure titles. }
\end{figure}

\begin{figure}
    \centering
    \includegraphics[width=\textwidth]{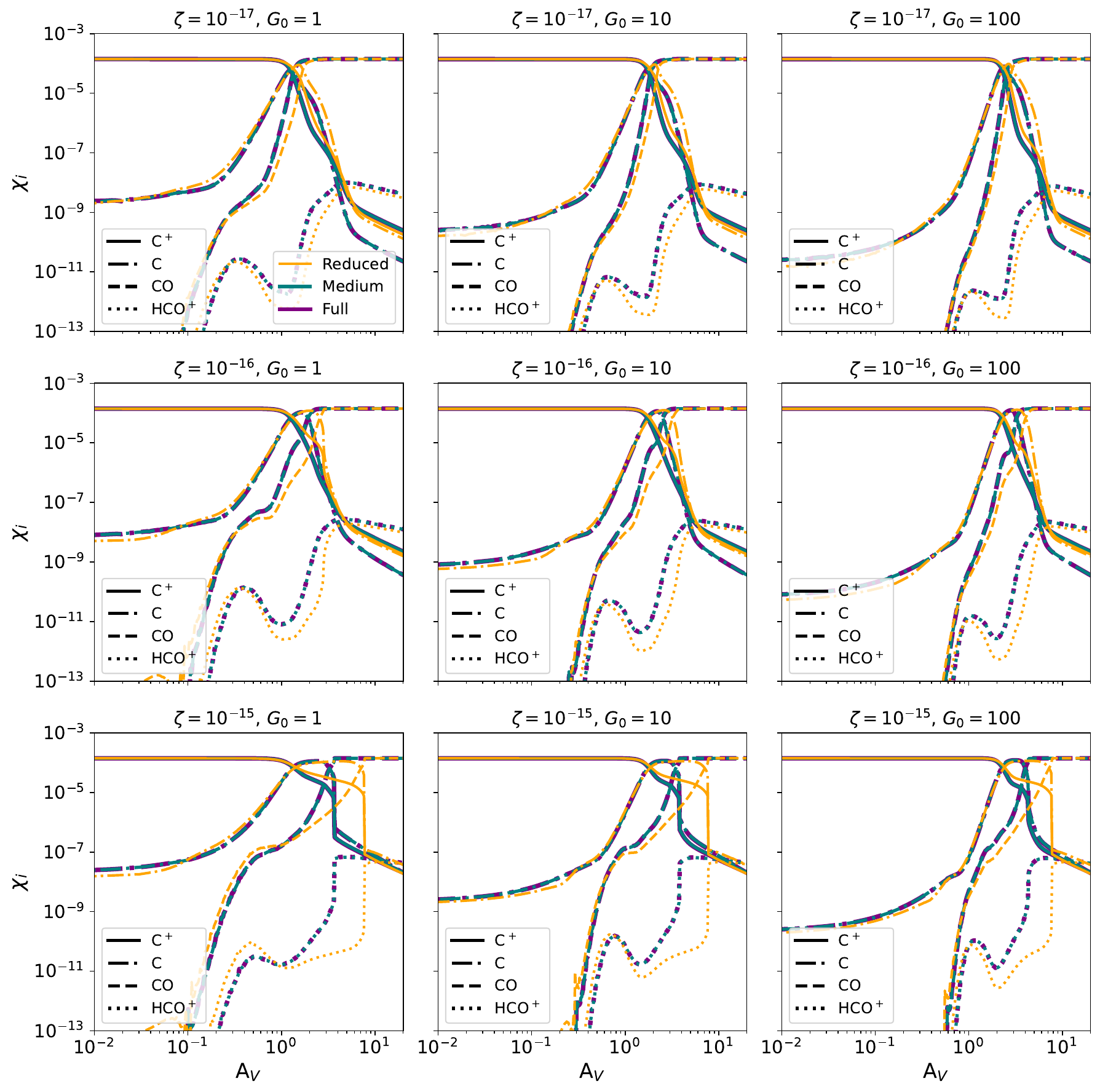}
    \caption{\label{fig:CmedBench}Same as Figure \ref{fig:HmedBench} but for the carbon cycle, \ce{C+}/C/CO/\ce{HCO+}.}
\end{figure}

\begin{figure}
    \centering
    \includegraphics[width=\textwidth]{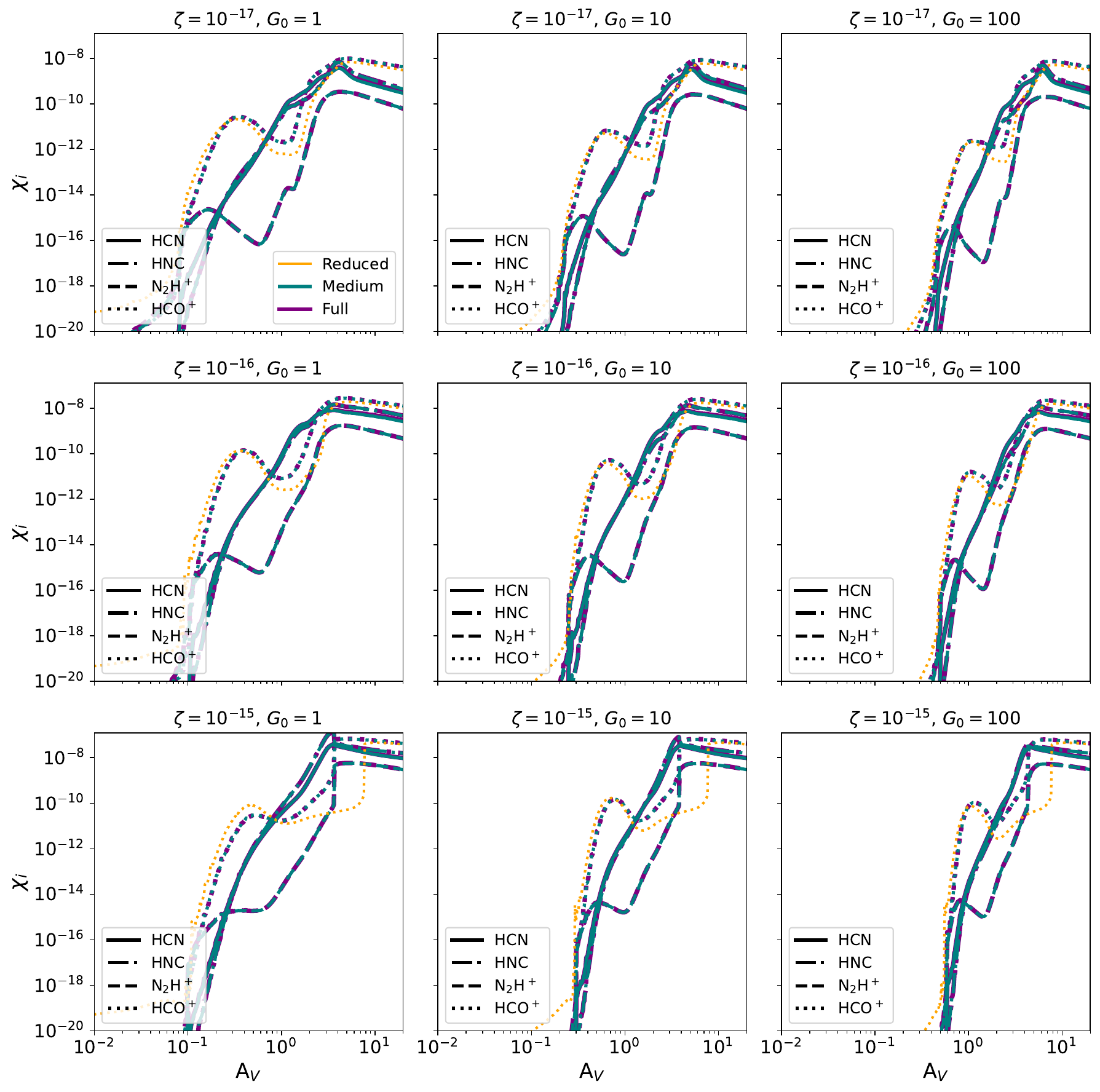}
    \caption{\label{fig:ismMedBench} Same as Figure \ref{fig:HmedBench} but for \ce{HCN}, \ce{HNC}, \ce{N2H+}, and \ce{HCO+}.}
\end{figure}

Figures \ref{fig:HmedBench}, \ref{fig:CmedBench}, and \ref{fig:ismMedBench} show the results of the steady-state benchmarking of the three different networks, reduced (orange), medium (teal), and full (purple). We find that, when compared to the full network, there is minimal difference for the species of concern. We have also performed this benchmark using constant density slabs in steady-state and similarly find no significant deviations for the species of interest. The network will be made available upon request and in the next major public release of {\sc 3d-pdr}. Since the networks are benchmarked in steady-state, it cannot be guaranteed that they have the same solution over temporal evolution.

\end{appendix}

\end{document}